\documentclass[sigconf]{acmart}
\usepackage{algorithm}
\usepackage{algorithmic}
\usepackage{enumitem}
\usepackage{graphicx}
\usepackage{subfig}

\usepackage{xcolor}
\definecolor{REVISEcolor}{HTML}{0000FF}
\newcommand{\REVISE}[0]{}

\AtBeginDocument{%
  }

\copyrightyear{2025}
\acmYear{2025}
\setcopyright{acmlicensed}\acmConference[CHI '25]{CHI Conference on Human Factors in Computing Systems}{April 26-May 1, 2025}{Yokohama, Japan}
\acmBooktitle{CHI Conference on Human Factors in Computing Systems (CHI '25), April 26-May 1, 2025, Yokohama, Japan}
\acmDOI{10.1145/3706598.3714176}
\acmISBN{979-8-4007-1394-1/25/04}

\begin{document}

\title{CalliSense: An Interactive Educational Tool for Process-based Learning in Chinese Calligraphy}


\author{Xinya Gong} 
\orcid{0009-0005-6414-9351} 
\email{gongxinya123@gmail.com} 
\affiliation{%
  \institution{Department of Computer Science and Engineering, Southern University of Science and Technology} 
  \country{China} 
}
\authornote{Equal contribution.}

\author{Wenhui Tao} 
\orcid{0009-0003-2645-6444} 
\email{wenhui9703@gmail.com} 
\affiliation{%
  \institution{Department of Computer Science and Engineering, Southern University of Science and Technology} 
  \country{China} 
}
\authornotemark[1] 

\author{Yuxin Ma} 
\orcid{0000-0003-0484-668X} 
\email{mayx@sustech.edu.cn} 
\affiliation{%
  \institution{Department of Computer Science and Engineering, Southern University of Science and Technology} 
  \country{China} 
}
\authornote{Corresponding author.}

\renewcommand{\shortauthors}{Trovato et al.}

\begin{abstract}
  Process-based learning is crucial for the transmission of intangible cultural heritage, especially in complex arts like Chinese calligraphy, where mastering techniques cannot be achieved by merely observing the final work. To explore the challenges faced in calligraphy heritage transmission, we conducted semi-structured interviews (N=8) as a formative study. Our findings indicate that the lack of calligraphy instructors and tools makes it difficult for students to master brush techniques, and teachers struggle to convey the intricate details and rhythm of brushwork. To address this, we collaborated with calligraphy instructors to develop an educational tool that integrates writing process capture and visualization, showcasing the writing rhythm, hand force, and brush posture. Through empirical studies conducted in multiple teaching workshops, we evaluated the system's effectiveness with teachers (N=4) and students (N=12). The results show that the tool significantly enhances teaching efficiency and aids learners in better understanding brush techniques.
\end{abstract}


\begin{teaserfigure}
  \includegraphics[width=\textwidth]{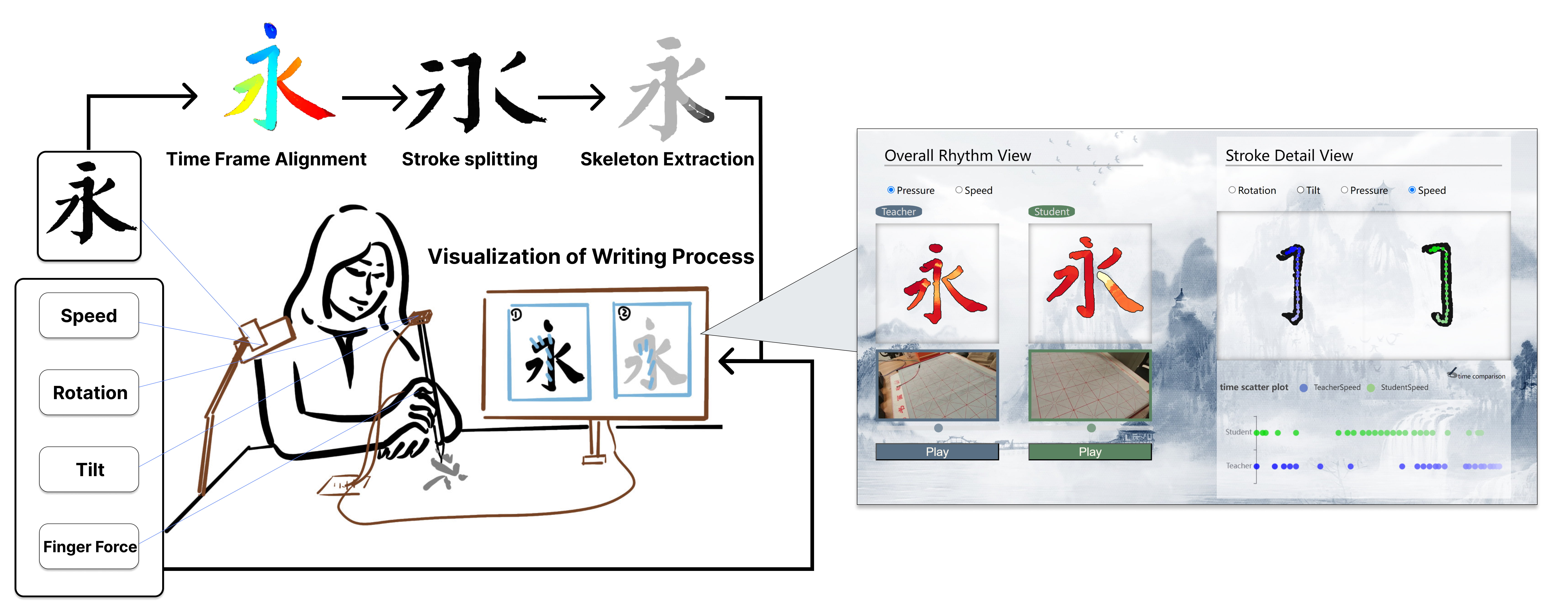}
  \caption{An overview the visual teaching assistant system for the entire calligraphy process begins with the user writing in a real scenario, where sensors and cameras capture the entire writing process. The captured strokes will then undergo a series of data processing steps, including temporal alignment, stroke segmentation, skeleton extraction, etc., and the captured writing parameters will correspond one-to-one with their own dots. CalliSense will visualize these parameters and has comparison features for both students and teachers, ultimately helping students enhance their awareness of brushwork and understand the details of stroke execution.}
  \label{fig:teaser}
\end{teaserfigure}


\maketitle
\section{Introduction}
The preservation and transmission of intangible cultural heritage (ICH) depend heavily on the effective teaching of skills and the documentation of the learning process. In this context, kinesthetic learning plays a critical role~\cite{TechnologyIntervention}, as mastering a skill involves not only understanding the final product but also perceiving and receiving feedback on the complex motor processes behind it~\cite{begel2004kinesthetic, magill2010motor}. In fields like artistic creation and traditional craftsmanship, while the final work is visible, the creation process often remains difficult to capture, making skill transmission particularly challenging~\cite{10.1145/3613904.3642205}. To address this, recent research has introduced a variety of technological solutions, such as gesture datasets are collected through motion segmentation in craft activities\cite{app10207325}, AR technology is used for interactive experiences\cite{app12199859}, and digital storytelling is conducted using virtual avatars\cite{https://doi.org/10.1111/j.1467-8535.2009.00991.x}. These technologies support multidimensional feedback that enhances learners' cognition and performance, which ultimately facilitates the effective transmission of ICH skills.

This paper takes the learning of brush techniques in Chinese calligraphy as a case study to investigate how capturing movement processes and providing visual feedback can support learning motor skills in the context of ICH. Our core research question is: What factors create barriers in the teaching and learning calligraphy brush techniques? Through a formative study, we found that beginner calligraphy students tend to focus too much on character shapes and often overlook the importance of brush control, which leads to a lack of understanding on proper brush techniques. This can result in the development of difficult-to-correct habits. Moreover, students often rely too heavily on external feedback and struggle to independently identify and correct mistakes. Since the dynamic process of brush control cannot be fully understood by observing static calligraphy, students also have difficulty capturing the details of the brush techniques during teacher demonstrations, especially the subtle timing of finger exertion, which is often neglected. Post-demonstration explanations are less effective due to students' fading memory of the brushstoke process, which further complicates the learning experience. Thus, capturing and clearly demonstrating the dynamic changes in brush control and finger force during writing is crucial for effective calligraphy learning.

To address these issues, a comprehensive solution, CalliSense, is proposed for capturing and visualizing the brushstroke process in Chinese calligraphy. Aiming for a widely applicable approach that allows experts and learners alike to contribute to the documentation of ICH motor skills, we opt for low-cost, accessible tools-specifically, smartphones, along with affordable pressure sensors and inertial sensors. These tools capture essential writing parameters such as brush posture, finger pressure, and writing speed, aligning them with the final written strokes. 
Ink deposition increments are used to mark time for tracking brush shape changes, while stroke points are filtered and interpolated to resolve brush occlusion and generate an accurate skeletal trajectory.
The captured process data is then visualized by incorporating traditional teaching routines, which also aligns with the typical practices of teachers and learners~\cite{ifenthaler2016learning}. The visualization design consists of three modules: character comparison, writing rhythm analysis, and stroke examination. Users are allowed to transit between modules seamlessly and receive a more holistic comprehension of the brushstroke process.
To validate the effectiveness of our system design, we organized four teaching workshops with 4 calligraphy instructors and 12 beginner students (i.e., students interested in calligraphy but with little experience). By comparing traditional teaching methods with those incorporating our CalliSense system, results showed that the system significantly aided students in understanding key aspects of brush techniques, especially those previously overlooked details.

In summary, our work makes the following contributions:

\begin{enumerate}
    \item A qualitative interviews with calligraphy teachers and learners of various skill levels to understand the challenges in the current transmission of calligraphy;
    \item CalliSense, a comprehensive system that includes the capturing of writing processes as well as a visualization interface to reveal brush control in Chinese calligraphy;
    \item A user study to assess the practicality and effectiveness of the system;
    \item Innovative design insights for using digital technology to facilitate the transmission of motor skills in intangible cultural heritage.
\end{enumerate}

\section{Related Work}
\subsection{Motion Capture and Analysis in Intangible Cultural Heritage}
The motion processes in intangible cultural heritage, such as calligraphy, dance, and traditional crafts, are important for skill transmission~\cite{bortolotto2007objects, lenzerini2011intangible, vecco2010definition}. With digital technologies including motion capture, sensors, and visual recognition, such processes can now be recorded accurately, which provides valuable data for research, education, and preservation.

Traditionally, intangible cultural heritage was passed down orally or documented in books. Early digital recordings focused on audio and video~\cite{pietrobruno2009cultural}, such as the preservation of Tiwi songs and dances in public archives~\cite{inbook}. Additionally, cultural soundscapes~\cite{samuels2010soundscapes} have been used to document and recreate the sound environments of cultures~\cite{noviandri2023cultural, Bartalucci_2020, doi:10.1080/13527258.2016.1138237}. Techniques like photogrammetry and optical motion capture have been employed to document heritage, as seen in the Quanzhou Chest-Clapping Dance~\cite{kirchhofer2011cultural, chen2014}. Modern approaches, such as wearable devices~\cite{protopapadakis2020digitizing}, multi-camera LiDAR\cite{caterina_balletti__2023}, and multimodal platforms like i-Treasures\cite{8255779} and CHROMATA, enable detailed modeling, analysis, and immersive virtual experiences\cite{9480948, selmanovic2018vr}, offering innovative ways to preserve and display intangible cultural heritage.

\begin{figure}[t!]
  \centering
  \includegraphics[width=\columnwidth]{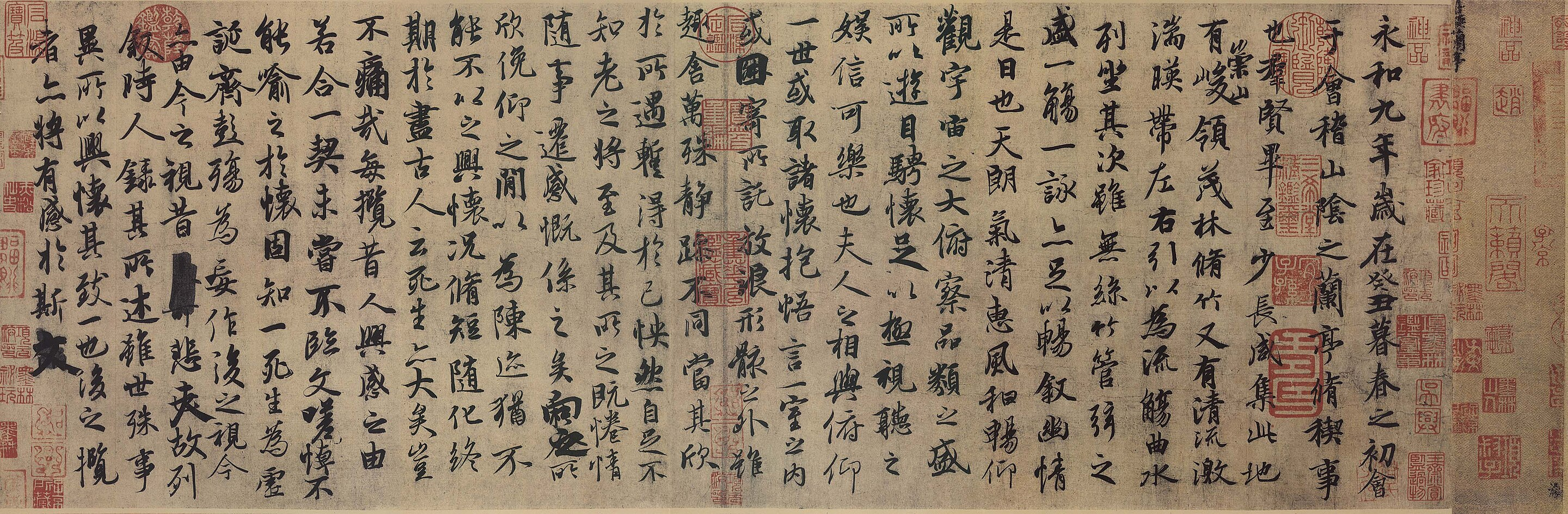}
  \caption{Reproduction of the famous Chinese stele inscription \textit{Lantingji Xu} from {Wikipedia}\protect\footnotemark}
  \label{fig:lantingji_xu}
\end{figure}

In the field of Chinese calligraphy, various technologies have been introduced to aid learning and replicate traditional writing processes. For example, although they differ from the actual writing experience, plotters and robotic arms attempt to replicate the movements of calligraphers by simulating the reactive force of real handwriting~\cite{nishino2011calligraphy, 10.1145/3613904.3642792, 10.1145/3526114.3558657}. Additionally, remote haptic systems guide students in using calligraphy brushes via network-controlled devices to ensure high-quality tactile feedback~\cite{10.1145/1255047.1255063}. Specially designed brushes equipped with sensor technology are also used to record writing trajectories~\cite{10.1145/3029798.3038422, Matsumaru_2017jaciii} and convey some aspects of the motion~\cite{10.1145/3559400.3565595}. However, the authentic texture (Figure \ref{fig:lantingji_xu}) and feel of traditional brushes remain difficult to replicate with digital brushes. Furthermore, the digital exploration of calligraphy, including multi-target detection~\cite{10.1007/978-981-15-3867-4_27}, style classification~\cite{10.1007/978-3-031-41685-9_5}, and visual appreciation~\cite{zhang2023visual}, is enriching both the teaching and cultural preservation of calligraphy. 

Although existing technologies help preserve intangible cultural heritage such as dance and sculpture, the recording of the calligraphy writing process remains limited and often focuses on imitation rather than authentic documentation. Current robotic arms and plotters attempt to replicate the writing process, but they fall short of capturing the true creative act. Additionally, modern technology primarily focuses on recording the character form, with little research on how brush posture and hand pressure affect the outcome.

For motion-based intangible heritage, most studies focus on body posture, with less attention paid to the control and movement of tools. As the medium connecting the hand to the paper, the brush's role deserves further exploration. CalliSense aims to document the brush's movements during natural writing and explore the relationship between motion and line formation.

\subsection{Visual Feedback Through Motion Analysis in Kinesthetic Learning}
\footnotetext{\url{https://zh.wikipedia.org/wiki/File:LantingXu.jpg\#file}}
Proprioceptive demonstration is a practice that enhances an individual's perception of their own position, movement, and posture through bodily actions and sensory feedback~\cite{tuthill2018proprioception, 10.1080/00222895.1974.10734977, winter2022effectiveness}. Individuals can experience specific movement trajectories guided by external forces or actively execute actions relying on their own sensory feedback~\cite{wong2012can, FARRER2003609}. Richard A. Magill's concept of kinesthetic feedback emphasizes the natural perception of body position, speed, and force, helping to adjust and optimize movements for more precise control, supporting the effectiveness of learning calligraphy skills through bodily perception in a natural writing state~\cite{magill2010motor}. 

In the field of human-computer interaction, kinesthetic feedback relies on efficient motion capture. In skill motion capture involving tools, such as archery, sensor technologies are widely applied. For example, Zhao et al. used wearable devices and accelerometer-based methods to analyze archers' "release" actions~\cite{zhao2016archery}. Similarly, Phang et al. utilized inertial measurement units (IMU) and fast Fourier transform (FFT) methods to examine the micro-movements of the bow~\cite{phang2024archery}. In golf~\cite{king2008wireless, fitzpatrick2010validation}, Nam et al. applied inertial sensors and stereo camera techniques to track the motion trajectory of golf clubs, demonstrating the effectiveness of these technologies in capturing detailed tool dynamics for skill enhancement~\cite{nam2013golf}.

In terms of feedback, numerous studies have demonstrated that providing rich, visual feedback through technology can significantly improve skill training. For example, in areas such as dance, painting, and musical instrument performance, techniques like beat comparison~\cite{10.1145/3274247.3274514}, shadow generation~\cite{10.1145/2010324.1964922}, hand motion capture~\cite{10.1145/3544549.3585838}, and the use of virtual avatars~\cite{10.1007/978-3-030-36126-6_7} and projections~\cite{5557840} for learning and feedback have proven effective. These methods allow learners to more intuitively understand and adjust their movements. Additionally, AR real-time feedback, visual cues, and multimodal feedback are widely used in fitness and music instruction~\cite{10.1145/3411764.3445649, 10.1145/3491102.3517735, 10.1145/3411764.3445595}. In golf, similar to Chinese calligraphy, rigid body devices are used to study various feedback mechanisms, including body angle visualization~\cite{10.1145/1878083.1878098}, ground shadows~\cite{10.1145/3305367.3327993}, sound cues~\cite{10.1145/3305367.3327993, 10.1145/3281505.3281604}, posture comparison waveforms~\cite{10.1145/3476124.3488645}, and multimodal feedback comparisons~\cite{10.1145/3427332}.

In Chinese calligraphy research, visual enhancement techniques have proven effective in improving character training. For instance, augmented reality overlays digital information onto the real world, providing real-time guidance and feedback for calligraphy learning, thus enhancing efficiency and experience~\cite{10.1145/1935701.1935769}. Interactive projection systems also help learners practice character shapes effectively by projecting calligraphy fonts and stroke sequences in real-time~\cite{9122337}. Other studies use virtual paper and digital brushes to offer real-time feedback and animation demonstrations~\cite{sym11091071}.
Current research largely focuses on the correctness of the final result, with limited attention to the detailed process of brush handling. Even when some studies attempt to correct the writer's posture, they often emphasize overall body posture rather than the fine control of the hand over the brush~\cite{10.1145/1935701.1935769}, or simply stress following the correct stroke path~\cite{10.1145/3029798.3038422}. Other studies have explored brush pressure and movement speed~\cite{10.1145/3377325.3377534}, but have not fully analyzed how these parameters affect the quality of the strokes, nor do they offer specific guidance for students practicing key techniques. Additionally, no human-computer interaction technology currently records and replays the detailed brushwork of students' writing processes in their natural state. In the next section, we will delve into the challenges of process learning in calligraphy and demonstrate how the CalliSense system addresses these issues.

\section{Formative Study}
The key to learning Chinese calligraphy lies in mastering fundamental techniques and maintaining consistent practice. The learning process begins with adopting the correct brush-holding posture, understanding basic strokes, and grasping the structure of characters\cite{zhang2023bringing, shi2019chinese}. Through repeated practice, students can develop muscle memory, improve stroke accuracy, and master the rhythm and fluidity of writing, which are crucial for the expressiveness of calligraphy. Incorporating cultural elements can make the learning process more engaging, while timely feedback and encouragement help students progress and express themselves\cite{hue2010aestheticism}. As students delve deeper into their studies, they can explore their own creativity within the traditional framework and enhance their understanding of calligraphy. This practice blends brushwork and ink control, allowing the mastery of angles, pressure, and speed to produce lines rich in personality and emotion\cite{chiang1974chinese}. To explore learning challenges in brush techniques, we conducted a formative study for design insights.

\subsection{Semi-structured Interviews}
Semi-structured interviews were conducted with eight calligraphy practitioners (four males, four females) to gain insights into their needs concerning the visualization of brushwork in calligraphy. The exploration focused on the following aspects:
\begin{enumerate}
    \item The challenges calligraphy practitioners face in their practice, particularly in brush handling, and which of these challenges are the most difficult to overcome.
    \item Identifying design opportunities for CalliSense to address these challenges.
\end{enumerate}

Our tool is designed for calligraphy learners that include practitioners of varying skill levels, ranging from beginners to calligraphy experts. Through the multi-level interviews, we aimed to understand the obstacles learners face in their practice and to gather insights from teachers on the common difficulties students encounter and the challenges they face in instruction. Additionally, it is believed that individuals who have planned to learn calligraphy but ultimately quit due to various reasons also offer valuable perspectives for the interviews. This group can reveal key barriers to learning traditional culture and help us identify and address potential issues.

The interviewees included two calligraphy experts (A1 and A2), who have been deeply involved in the field for 15 and 35 years, respectively, and are well-regarded in the community. Both experts have extensive experience in practicing and teaching calligraphy and will continue to serve as consultants, providing ongoing insights for our design process. The remaining six interviewees were volunteers recruited online. Four of them are calligraphy practitioners (B1–B4), representing different age groups  (mean = 33, SD = 13.4) and having practiced for over a year, with a basic understanding of calligraphy techniques. The remaining two interviewees (C1 and C2) are individuals who previously practiced calligraphy but have not continued for nearly a year. Though their experience was brief, they offer a valuable beginner's perspective.

The interviews began by asking the three groups about the difficulties they encountered in calligraphy practice and the methods they used to overcome them. This was followed by a discussion on the key aspects they focused on during practice. For the two calligraphy experts, we additionally explored their teaching experiences to further investigate the most challenging issues students usually meet in the learning process. 
 
\subsection{Analysis and Results}
Our semi-structured interviews were conducted over the phone, with each interview lasting approximately 30 minutes and recorded for later transcription. To analyze the interview data, we employed thematic analysis. First, the co-authors read through the transcripts to gain an overall understanding. After familiarizing themselves with the data, they independently performed open coding. Upon completing the coding, the co-authors shared their results, discussed their interpretations of the data, and reached a consensus on the final coding outcomes. 

\subsubsection{The Lack of Awareness in Brushwork}

In this category, we have identified two issues:

\textbf{Beginners' Tendency to Overlook the Importance of Brushwork:}
For beginners, understanding correct brushwork means knowing what constitutes proper stroke techniques, brush pressure, and brush movements, as well as how to achieve these effects. Teachers believe that while mastering brush techniques requires long-term practice, having the right awareness beforehand is more crucial to avoid forming bad habits that are hard to correct (A1: ``If students realize the importance of brushwork and keep practicing, they will improve steadily. The real danger is when students mistakenly believe they are doing it correctly. That's when it's a problem.''). Using the correct methods makes learning more efficient, while the wrong approach could lead to frustration and even quitting (B3: ``Finding the right teacher is important. If the method is correct, it becomes easy to apply knowledge in new situations. Otherwise, learning becomes difficult, and it's easy to give up.''). However, there is currently a shortage of calligraphy instructors, and learners often lack a full understanding of the importance of brushwork in the early stages, making it difficult to judge what constitutes a good piece of work (C1: ``When I practiced, I only focused on whether the characters looked right. I didn't pay attention to brushwork unless the teacher pointed it out.''). Therefore, it is crucial for calligraphy learners to recognize the key role that brushwork plays from the very beginning of their studies.

\textbf{Reliance on External Feedback}: Both calligraphy practitioners and teachers generally believe that beginners rely heavily on a teacher's guidance to avoid going astray (N=6) (A1: ``There's a widely shared, albeit somewhat `biased' saying in the calligraphy community: `Self-learning is tantamount to self-destruction.' ''). This is because learners can grasp brush techniques and variations in rhythm through a teacher's demonstration, something that is difficult to experience solely by following copybooks (N=7) (B2: ``When it comes to brushwork, I find a rolling stroke technique in cursive script particularly challenging. It requires the teacher's demonstration to fully understand it. It's hard to grasp from a copybook alone.''). Additionally, practitioners often find it difficult to correct their mistakes on their own during writing (N=5) (A2: ``I knew there was a problem, but I didn't know how to fix it. Learning helped me figure out how to make those corrections.'' B1: ``My hand trembles when I write, probably because I'm holding the brush too tightly, but I don't know why this happens.''). Even experienced calligraphy practitioners who have been practicing for years may need to attend specialized institutions and seek guidance from more experienced teachers to correct ingrained habits that are difficult to detect (A2: ``When I was practicing calligraphy, I always struggled to match the copybook style. It wasn't until I attended an institution that I realized my habits were off, and the learning process was quite painful.''). Therefore, it is essential for practitioners to receive feedback on their brushwork techniques during the writing process. 

\subsubsection{Difficult-to-Detect Brushwork Techniques} In addition, three aspects regarding learning specific techniques are identified:

\textbf{Difficulty in Observing Brush Techniques Through Strokes:} Unlike regular writing tools, the brush is challenging for beginners to master, and they may struggle with how to control it (N=7). (A2: ``Beginners find it hard to grasp the characteristics of the brush and don't know how to apply force.'') The static appearance of the characters makes it difficult to observe the brushwork process, which leads to situations where learners cannot find the connection between brush techniques and the resulting strokes when copying calligraphy models. They also don't know how to control the brush through hand movements, making it hard to achieve their writing goals (N=6). (B1: ``Strokes are really the result of brush-tip movements, but when I look at ancient calligraphy models, I can't see this. I don't know how to practice.'') Therefore, students need to be shown the brushwork process corresponding to specific calligraphy strokes.

\textbf{Demonstration Limitations:} Calligraphy instruction often involves numerous abstract terms, such as center stroke (zhongfeng), wrapping stroke (guofeng), and reverse stroke (nifeng). Students find it challenging to fully understand these terms through verbal explanations alone. To address this issue, teacher demonstrations are the most common and effective method. However, students have limited observational abilities and often miss critical details in the teacher's demonstrations (N=6). (A1: ``When discussing the concept of `center stroke', students might spread the brush bristles wide while writing a horizontal stroke, thinking this will create a thicker line. However, this approach does not align with the true principle of the center stroke.'')

During demonstrations, it is difficult for students to simultaneously analyze the stroke, hand movements, and the complex changes in the brush bristles. As a result, they still struggle to understand what kind of writing process corresponds to a specific calligraphy term. Therefore, when teaching brush techniques, it is essential to break down the movements of the brush in sufficient detail.

\textbf{Hand Force Cannot Be Observed:} Calligraphy writing relies on precise control of the brush by the hand, and learners typically rely on teacher demonstrations or videos to learn. However, not only is it difficult to capture the subtle movements of the brush handle, but the force applied by the hand to the brush is inherently hard to detect with the naked eye (N=5). (B1: ``When writing small characters, my hand often trembles. I'm not sure if I'm gripping the brush too tightly. The teacher didn't specifically mention this, and I can't figure it out.'' A1: ``I also realized that students were gripping the brush too tightly after teaching for a while. Now, I can roughly judge if they are applying force incorrectly based on the strokes, but younger teachers may not be able to do this.'')

Although teachers explain the brushwork details during demonstrations, they often do not cover all the key points and may not be aware of the specific aspects students are focusing on, making it challenging to address their learning needs. For instance, when writing long strokes, in order to achieve dynamic movement in the middle of the line and clean, crisp ends, the hand's force typically follows a pattern of tension—relaxation—tension. However, since the explanation often focuses on the tip of the brush, this technique is frequently overlooked by students. Therefore, the pressure applied by the hand to the brush needs to be demonstrated explicitly.

\subsubsection{Forgotten Writing Process: }In this category, two issues have been identified: 

\textbf{Forgetting Brushwork Details:} In traditional calligraphy instruction, to avoid interrupting students' writing, teachers typically provide feedback after the student has finished. However, by this time, the details of the writing process are often forgotten (B4: ``Sometimes I don't understand what the teacher is referring to, and I need to write it again to make sense of it''). On occasion, students even bring completed works to the teacher for critique. While experienced teachers can quickly identify issues, students, having forgotten their own writing process, struggle to connect the teacher's feedback with their performance at the time. As a result, they can only mechanically record the feedback and reflect on it later, missing the opportunity for real-time interaction with the teacher. Therefore, it is essential that students' writing process be more fully captured and reconstructed.

\textbf{Forgetting Writing Rhythm:} Writing rhythm is a critical element in calligraphy practice, as the distribution of hand force and writing speed throughout the character directly impacts the overall appearance of the Chinese character\cite{wang2024standards, kraus1991brushes}. In well-resourced teaching environments, instructors typically ask students to rewrite problematic strokes and provide feedback. However, even with rewrites, it is challenging for students to fully recreate the original writing rhythm. On the one hand, students may consciously adjust their natural writing state when being observed by the teacher. On the other hand, the distribution of force and speed during writing is complex, making it difficult to recall the overall rhythm from a macro perspective through rewriting alone. Therefore, students need an ``overview'' perspective to review and comprehend the complete writing rhythm.

\subsection{Design Consideration}
Based on the findings in the formative study, we identified five design considerations to build a system that supports designers' reference recombination process during early-stage ideation: 

\textbf{DC1: Capture the Complete Writing Process} 
To accurately reconstruct the writing motion, the system should capture the entire writing process of a character, including detailed brush techniques, ensuring that the dynamic changes of each stroke and brush movement are recorded.

\textbf{DC2: Correlate Strokes with Brush Techniques}
Although the quality of the lines can be directly assessed through the writing results, the specific brush techniques used to create these lines are equally important. The system should be capable of precisely pinpointing specific segments of the strokes and reviewing corresponding brush details, such as brush posture and finger pressure. At the same time, it is essential to ensure that the writing process aligns with the final strokes for accurate analysis.

\textbf{DC3: Review Overall Writing Rhythm}
Writing rhythm influences the internal contrast and variation within a character. Therefore, the system should support reviewing the overall rhythm after writing is completed, allowing users to observe the variations in hand force and writing speed throughout the character.

\begin{table*}
\caption{Brush Measurement Parameters and Their Impact in Calligraphy}
\label{tab:calligraphy-measurements}
\setlength{\tabcolsep}{4pt}   
\renewcommand{\arraystretch}{1.5}  
\begin{tabular}{p{3cm} p{4cm} p{9cm}} 
\hline
\textbf{Measurement Object}   & \textbf{Measurement Parameter}  & \textbf{Impact and Significance in Calligraphy} \\ \hline
\textbf{Brush Handle}  
    & Tilt  
    & Affects the friction between the brush and the paper, which in turn alters the strength and expressiveness of the strokes. \\ \cline{2-3} 
    & Rotation  
    & Influences the position and organization of the brush tip, thereby impacting the texture of the strokes. \\ \cline{2-3} 
    & Speed  
    & Affects the contact time between the brush and the paper, altering ink absorption and subsequently affecting stroke thickness and ink intensity. \\ \hline
\textbf{Fingers}  
    &  
    & Affects how the brush hairs interact with the paper, thereby influencing the texture of the strokes. \\ \hline
\textbf{Brush Tip}  
    &  
    & Affects the force applied through the brush to the paper, leading to variations in the texture of the strokes. \\ \hline
\end{tabular}
\vspace{10pt}
\end{table*}

\begin{figure*}[t!]
  \centering
  \includegraphics[width=\textwidth]{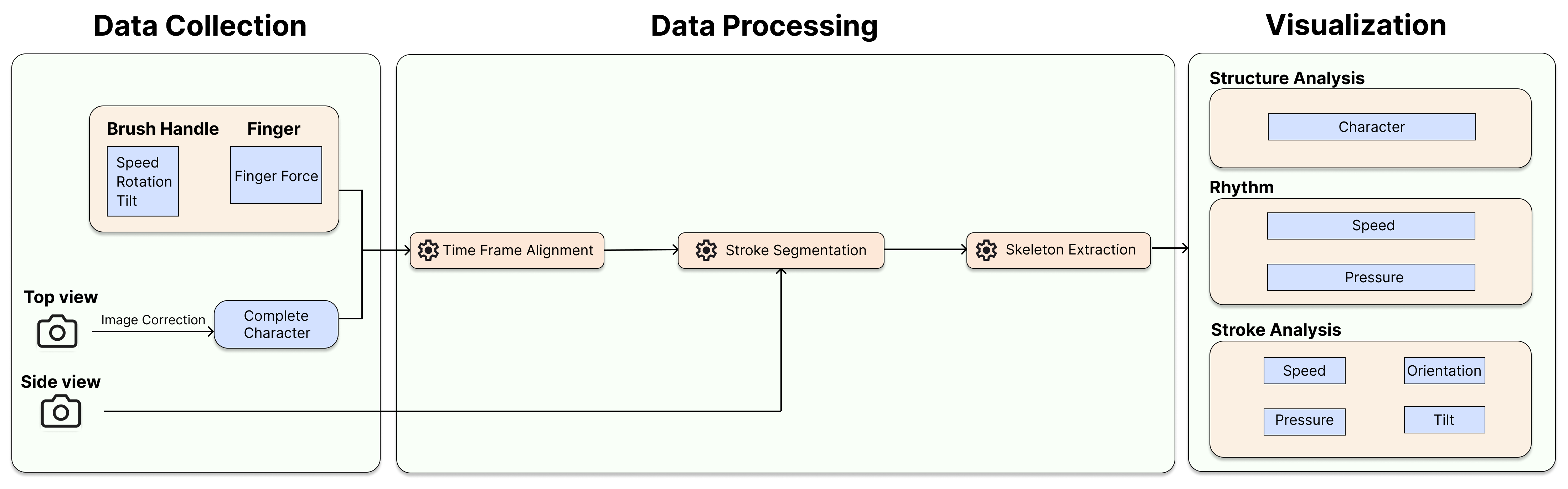}
  \caption{Overview of the work: The entire process from data collection and processing to visualization}
  \label{fig:overview}
\end{figure*}

\textbf{DC4: Examine Brush Tip Dynamics}
The brush tip is a critical factor that directly affects stroke quality. The system should provide a clear visualization of the brush tip's behavior during writing.

\textbf{DC5: Identify Brush Technique Errors Through Comparison}
The system should enable users to compare their brush techniques with those of the instructor, helping to quickly identify errors in brush posture, hand force, writing rhythm, and other aspects, while providing effective feedback for improvement.

Based on these five design principles, we propose the CalliSense system, an interactive tool for visualizing the calligraphy writing process. The system consists of two main components: (1) a camera and sensor suite for capturing the writing process, and (2) a web interface for visualizing the brushstroke process and allowing user control.

\section{Data Collection and Processing}
Considering the need for a complete record of both the strokes and the brushstoke process \textbf{(DC1, DC4)}, we design a data collection methodology that utilizes cameras to capture the brushstoke process and sensors to collect brush posture and finger pressure \textbf{(DC2)}. Algorithmic techniques are then applied to align the brush parameters with the stroke positions \textbf{(DC2)}. After discussions with experts, the specific collection parameters and their significance were finalized, as shown in the Table \ref{tab:calligraphy-measurements}. The overall data workflow of the system is illustrated in Figure \ref{fig:overview}, with the detailed design and corresponding descriptions discussed in the following sections.

\subsection{Image Data Collection}
In this step, the goal is to capture raw video footage of the writing process, laying the groundwork for subsequent data processing. 
\subsubsection{Equipment Overview}
\begin{figure}[t!]
    \centering
    \includegraphics[width=0.4\textwidth]{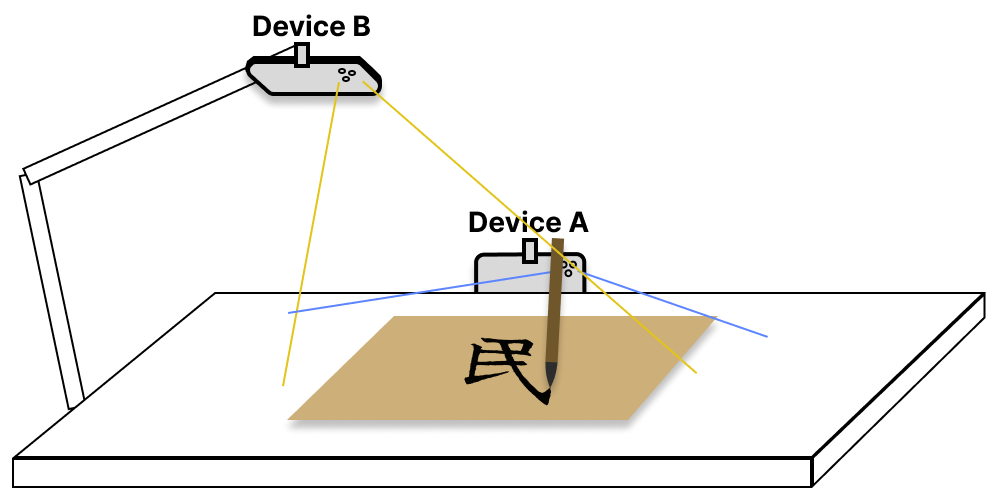}
    \caption{Capture the brushstoke process using two smartphones from different angles: Device A and Device B}
    \label{fig:device A and B}
\end{figure}

To minimize user reliance on specific equipment, we use widely available smartphones as the primary data collection tool. We have set up two recording devices, referred to as \textbf{Device A} and \textbf{Device B} (Figure \ref{fig:device A and B}). \textbf{Device A} is placed parallel to the table surface to monitor the contact between the brush tip and the paper, detecting writing motions. \textbf{Device B} is fixed above the table, covering the entire writing area to capture the full brushstroke process.

\begin{figure}[t!]
    \centering
    \includegraphics[width=0.48\textwidth]{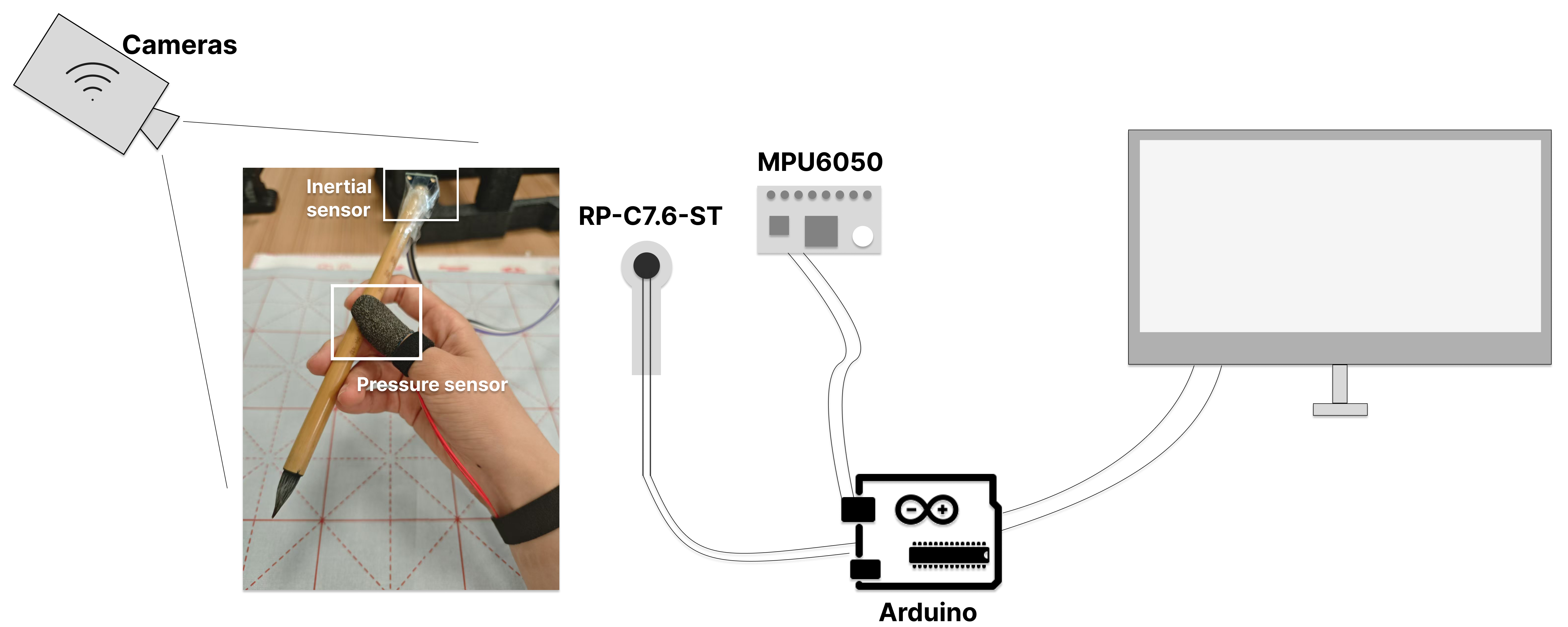}
    \caption{The data collection setup includes pressure sensors, inertial sensors, and cameras. The sensors are connected via Arduino to capture data, which is then synchronized and processed on the computer.}
    \label{fig:whole system}
\end{figure}

\subsubsection{Image Preprocessing}
\begin{figure}[htbp]
    \centering
    \includegraphics[width=0.4\textwidth]{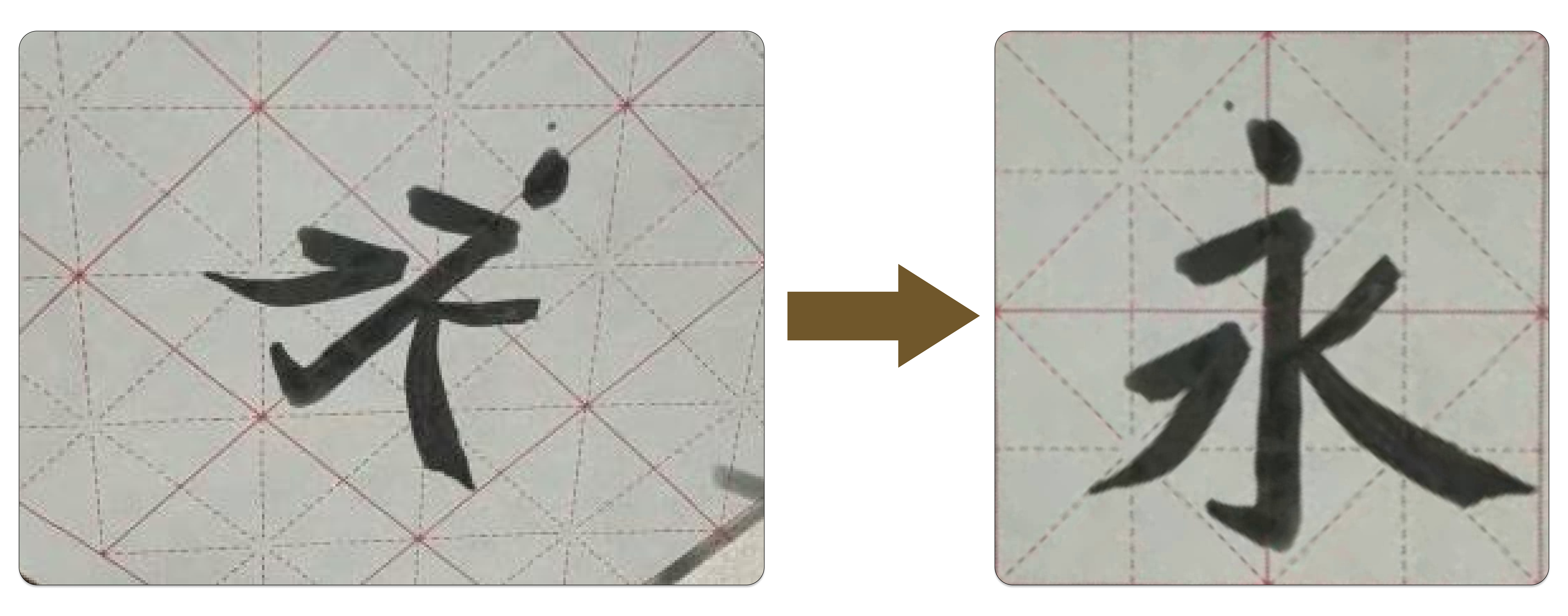}
    \caption{Stroke Images Before and After Correction}
    \label{fig:view_correct}
\end{figure}

After \textbf{Device B} recording the whole brushstoke process, the videos are sent to a data preprocessing stage. To correct the perspective distortion captured by the overhead camera, we use the perspective transformation algorithm from the OpenCV library~\cite{opencv_library}. This algorithm corrects image distortion by defining source and destination points, calculating a perspective transformation matrix, and remapping the image to a top-down view (Figure \ref{fig:view_correct}). This step ensures that the writing trajectory is presented at the correct scale and angle in the video, accurately restoring the original writing result during subsequent stroke visualization \textbf{(DC1)}. 

\subsection{Sensor Data Collection}
In this step, the inertial sensor and pressure sensor are used to capture the brush's posture and the hand's applied force, respectively (Figure \ref{fig:whole system}). 
 
\subsubsection{Brush posture collection}
The MPU6050 inertial sensor was used to measure the brush's tilt and rotation, and it was attached to the brush head. The sensor's built-in Digital Motion Processor (DMP) module can directly process data from the accelerometer and gyroscope to calculate the object's orientation. This not only reduces the computational load on external processors but also provides real-time orientation data~\cite{widagdo2017limb}. 

\subsubsection{Finger force collection}
Previous research inferred hand force by detecting arm muscle activity~\cite{10269740, 10340786}, but we required more precise finger force data due to the critical role of the fingers in writing. Therefore, we selected a circular resistive pressure sensor (short tail RP-C7.6-ST-LF2), which offers high sensitivity, small size, quick response, and low cost. Since the brushstoke process requires rotating the brush, we opted to attach the sensors to the fingers rather than directly to the brush shaft. This allows users to practice with any standard brush without the need for modifications. We initially considered wrapping a strip-shaped sensor around the shaft, but due to the slender nature of the brush handle, the excessive curvature of the sensor could lead to measurement errors.
\begin{figure}[t!]
  \centering
  \includegraphics[width=0.4\textwidth]{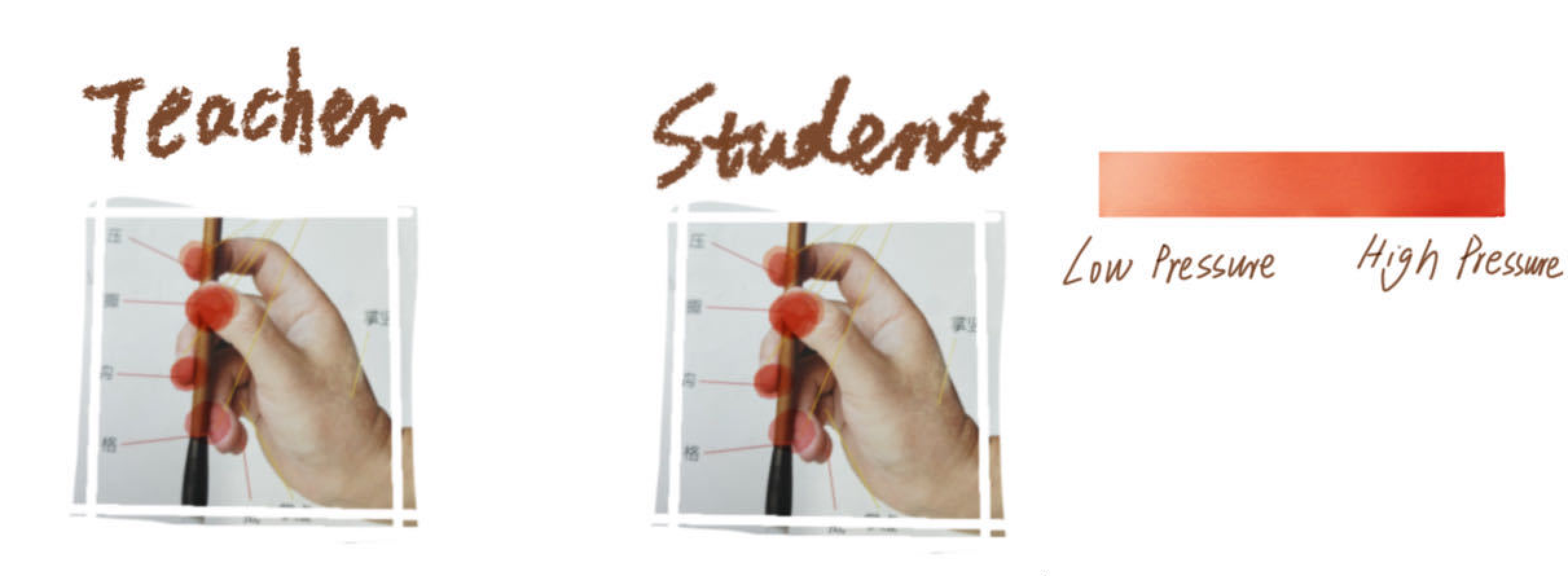}
  \caption{Design sketch: the pressure each finger applied to the brush}
  \label{fig: finger}
\end{figure}
Based on the posture of holding the brush handle, we attached four sensors to the thumb, index finger, middle finger, and ring finger using finger sleeves to measure the pressure each finger applied to the brush handle (Figure~\ref{fig: finger}). However, during subsequent iterations, expert suggested that the force exerted by the hand is primarily applied to the brush shaft, indirectly altering the contact between the brush tip and the paper, thus affecting the quality of the strokes. Such details are usually not emphasized in typical teaching settings, where only the overall force application needs to be monitored. Additionally, monitoring the force of multiple fingers might confuse beginners, leading them to believe they need to intentionally replicate different force applications from all four fingers, thereby creating unnecessary learning challenges.

As a result, we simplified the original four sensors to a single one, placed on the thumb. The thumb, being positioned opposite the other three fingers, serves as the primary point of force during writing and can largely represent the overall force exerted by the hand on the brush.

\subsection{Time Alignment and Skeleton Extraction}
 \begin{figure*}[htbp]
    \centering
    \includegraphics[width=0.9\linewidth]{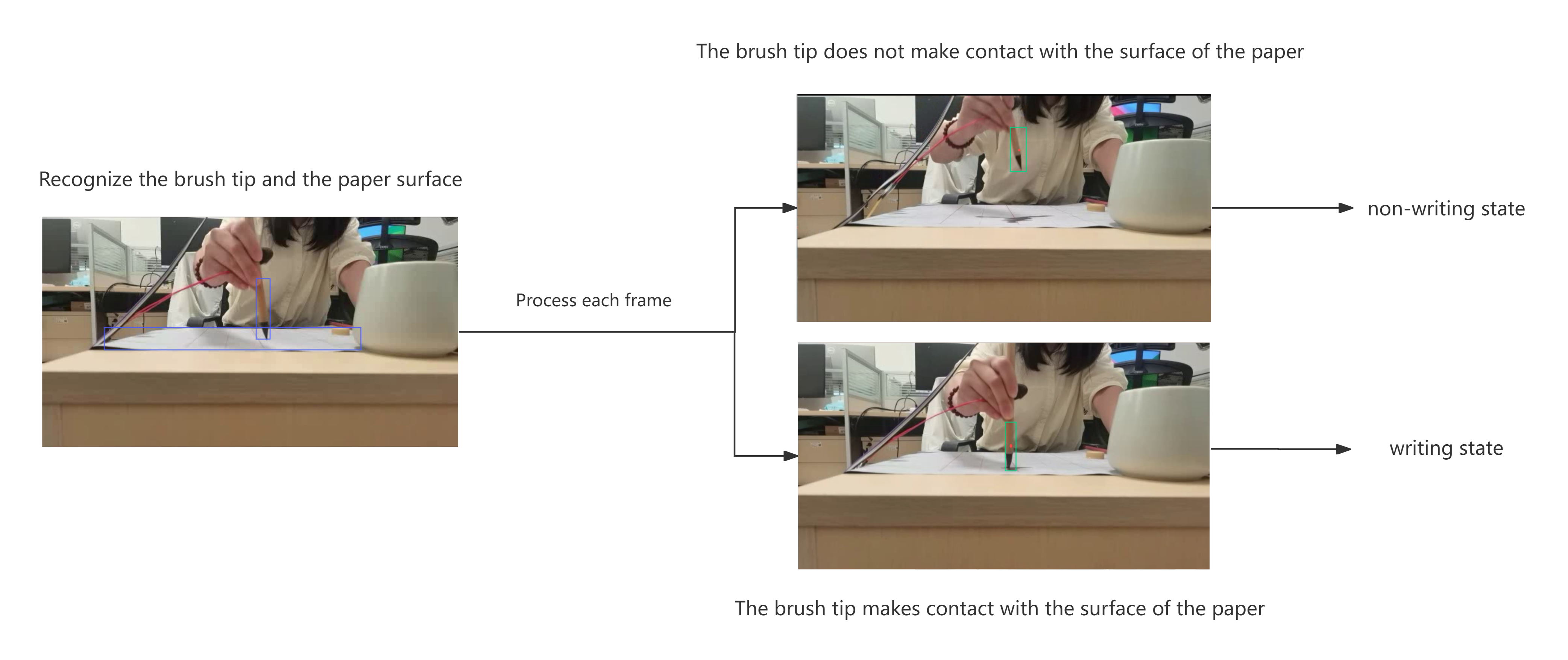}
    \caption{Determine if the brush is in writing mode based on the distance between the brush tip and the paper}
    \label{fig: Pen up and Down}
\end{figure*}
\subsubsection{Time Alignment}
\begin{figure}[H]
    \centering
    \includegraphics[width=0.6\linewidth]{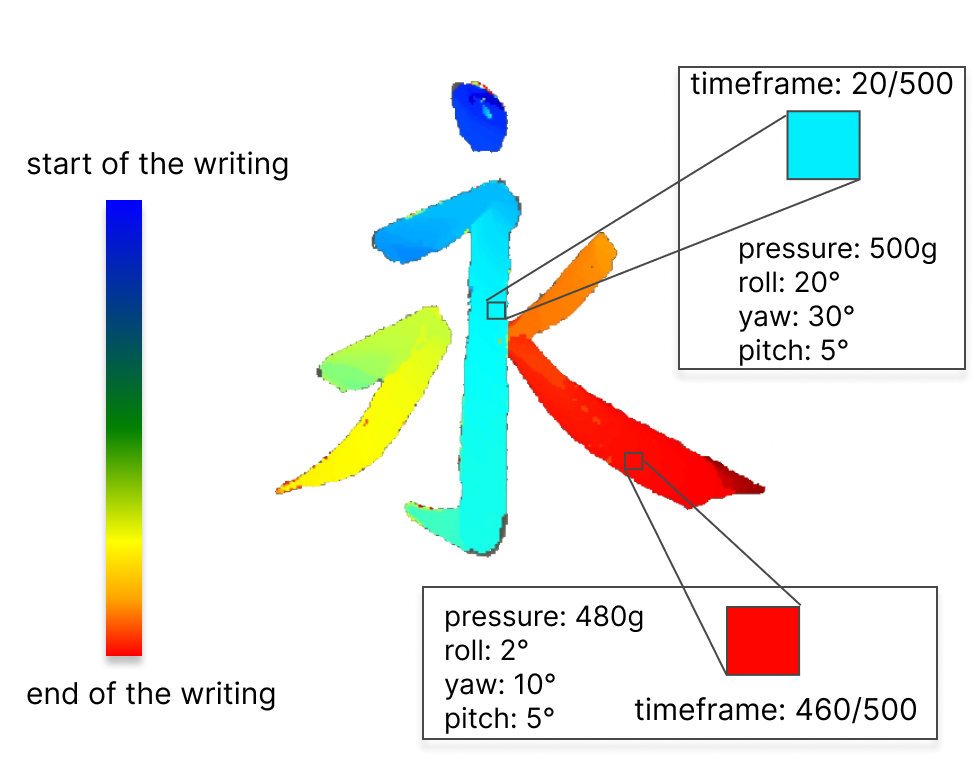}
    \caption{In the figure, each point of the handwriting is annotated with the corresponding timestamp information. The entire time frame is assumed to be 500 units, providing context for the timing of each point. }
    \label{fig: TimeAlignment}
\end{figure}

In order to match the collected strokes with the sensor data, we need to assign a timestamp to each stroke point and align the stroke's time labels with the sensor data. The most straightforward approach is to record the time when the ink appears, thus corresponding the ink with the timeline, as shown in Figure~\ref{fig: TimeAlignment}.

\begin{figure}[!htbp]
\centering
    \subfloat[A Chinese character can have different stroke segmentation methods.]{ \includegraphics[width=0.8\linewidth]{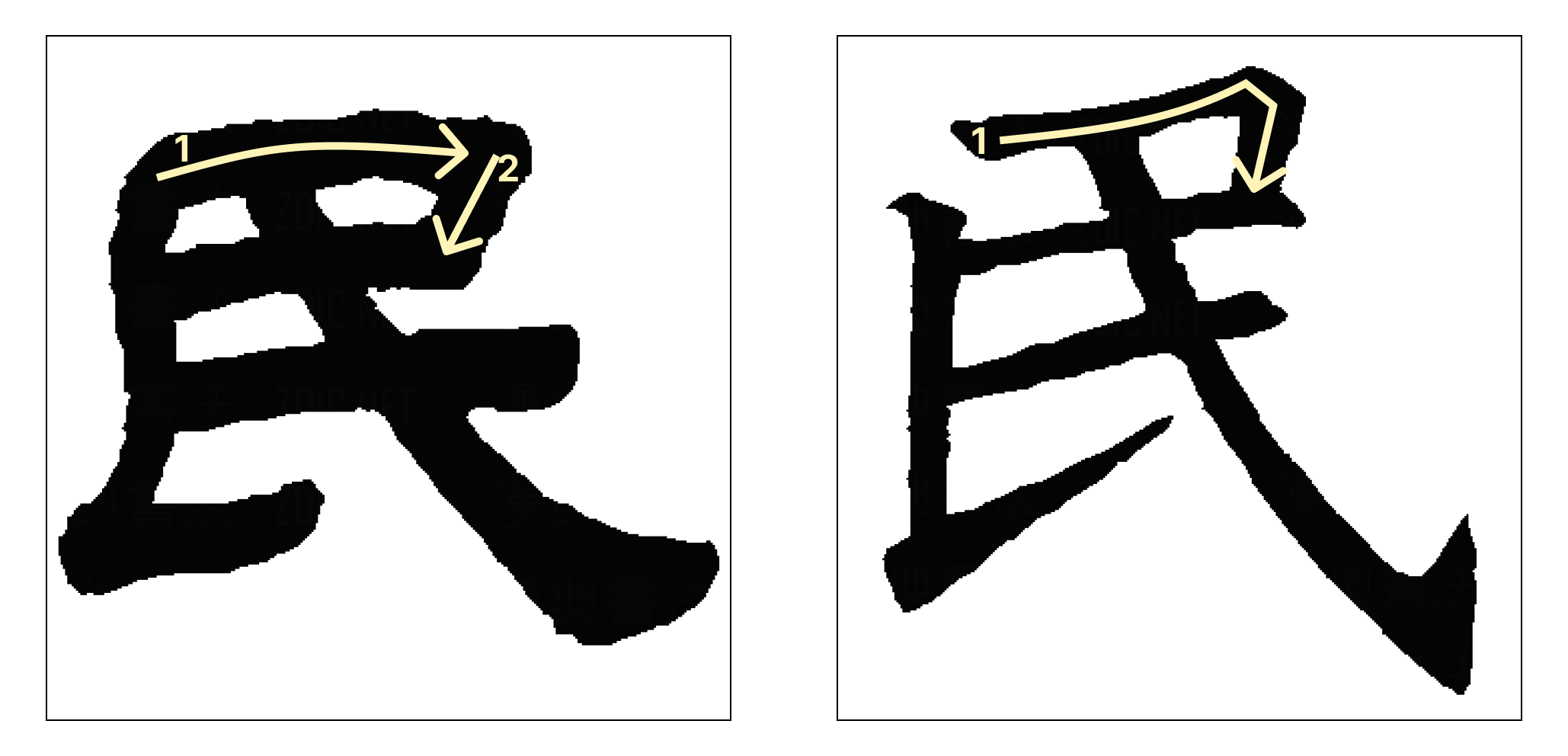}}\\
    \subfloat[The character ``Yong'' is segmented into different individual strokes.]{\includegraphics[width=0.9\linewidth]{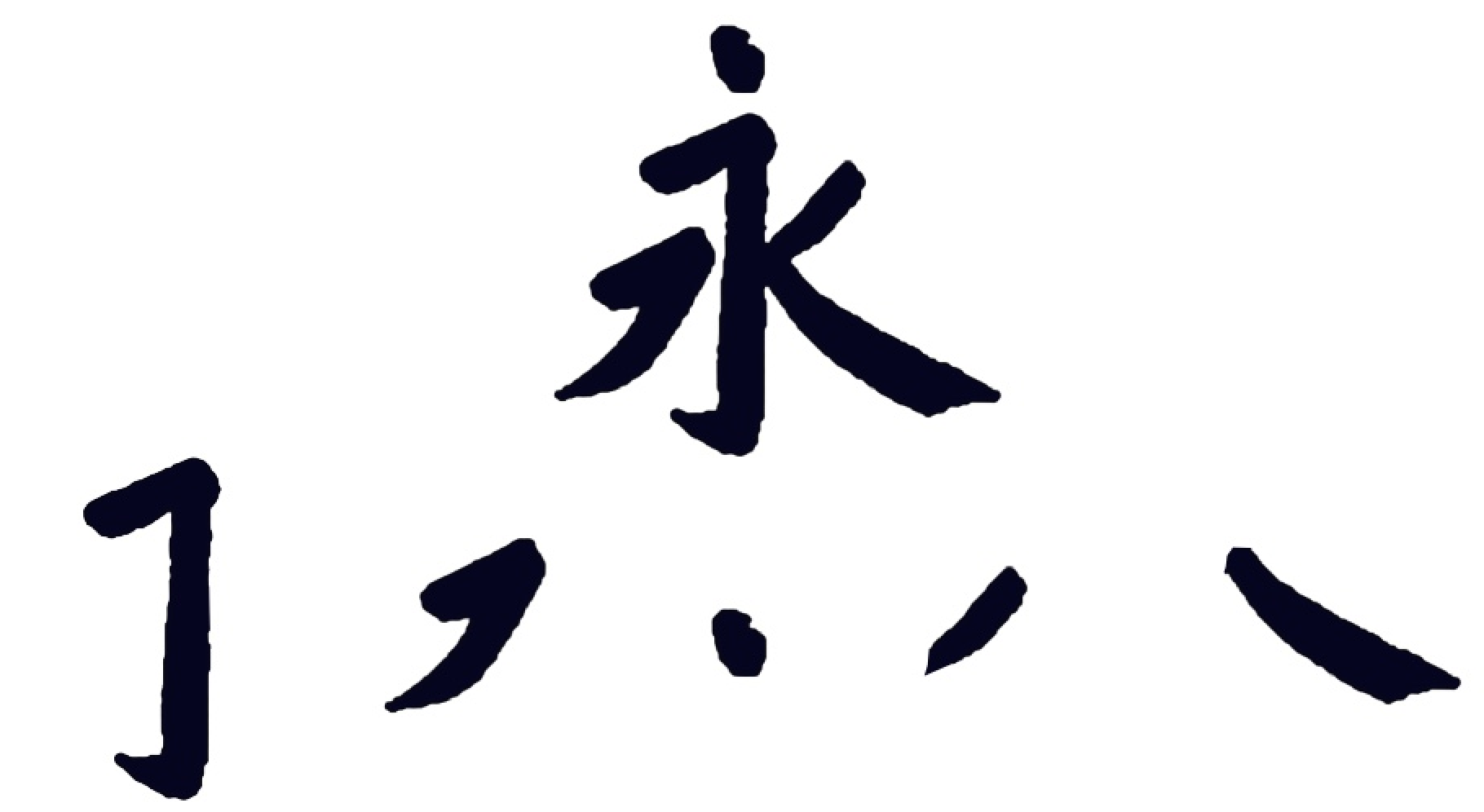}}
    \caption{Segmentation of Calligraphy Strokes.}
    \label{fig: image_of_min_charactor}
\end{figure}

Once the stroke's timestamps are assigned, we perform stroke segmentation to analyze the writing process of each stroke in detail \textbf{(DC2)}. Traditional stroke segmentation methods typically handle flat Chinese character images~\cite{10.1145/3548608.3559239, YAO2004631}. However, the rules for segmenting strokes vary between different calligraphy styles. For example, in the character ``Min'', the horizontal and vertical strokes in Clerical Script are made with two separate strokes, while in Regular Script, they are completed in one stroke (Figure \ref{fig: image_of_min_charactor}a).

\begin{figure}[!htbp]
    \centering
    \includegraphics[width=0.8\linewidth]{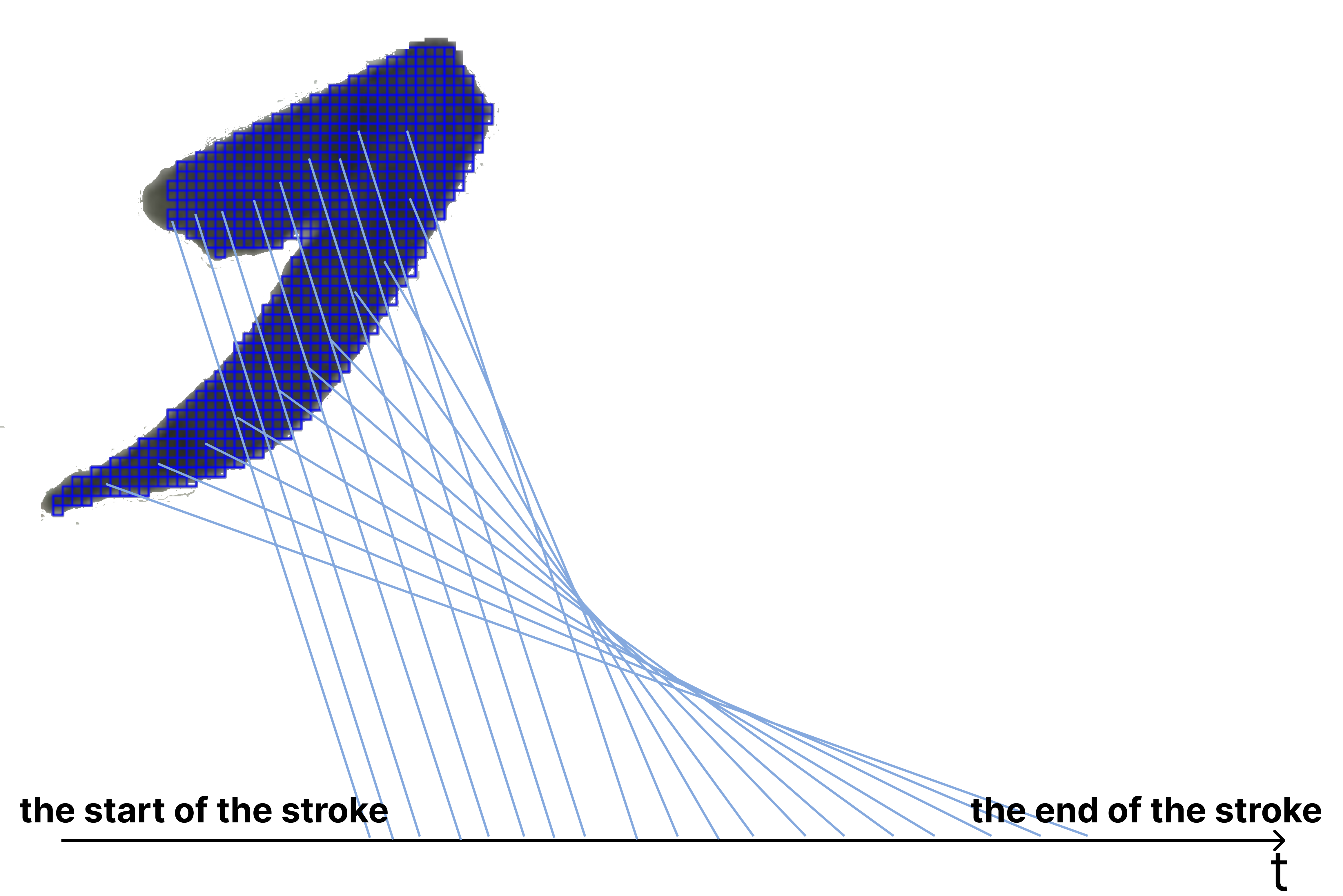}
    \caption{Pixel points are assigned to brush lift/drop intervals, isolating a stroke in the character ``Yong''.}
    \label{fig: strokepointsgroup}
\end{figure}

To enhance the system's adaptability to different calligraphy styles, we adopt an image-based method, segmenting strokes based on the start and end of each brushstroke in the writing process. Specifically, a side-view camera (Device A) combined with OpenCV's CSRT-based tracker is used to detect the contact between the brush tip and the paper (Figure~\ref{fig: Pen up and Down}). The time points of each brush's lifting and landing, combined with the stroke's time data, allow for stroke-by-stroke classification while maintaining the stroke order (Figure~\ref{fig: strokepointsgroup}). Taking the character "Yong" as an example, it is divided into five strokes based on the state of contact between the brush and the paper (Figure \ref{fig: image_of_min_charactor}b).




\subsubsection{Skeleton Extraction}
\begin{figure}[H]
    \centering
    \includegraphics[width=0.4\textwidth]{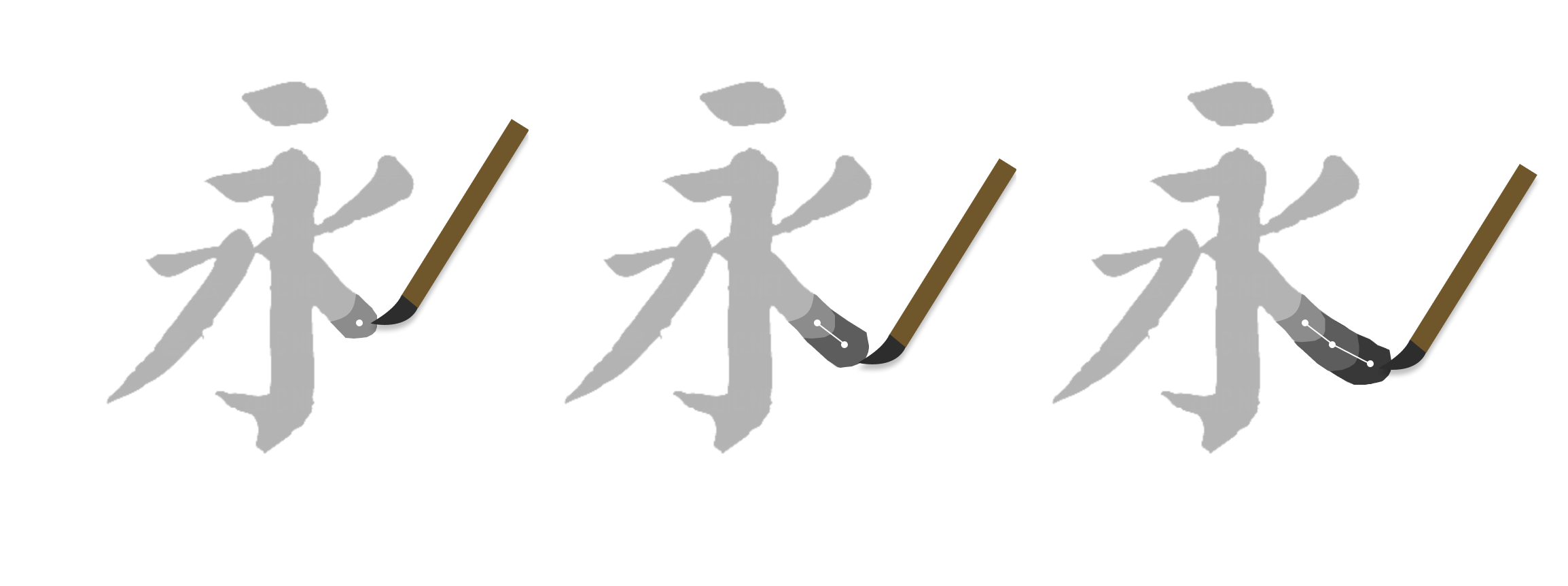}
    \caption{By tracking the added ink traces, time frames are assigned to different positions of the character. The pixel positions from each frame are averaged to generate central axis points, which are then connected to form the character's central axis line.}
    \label{fig: Ink increment}
\end{figure}

\begin{figure}[H]
    \centering
    \includegraphics[width=0.9\linewidth]{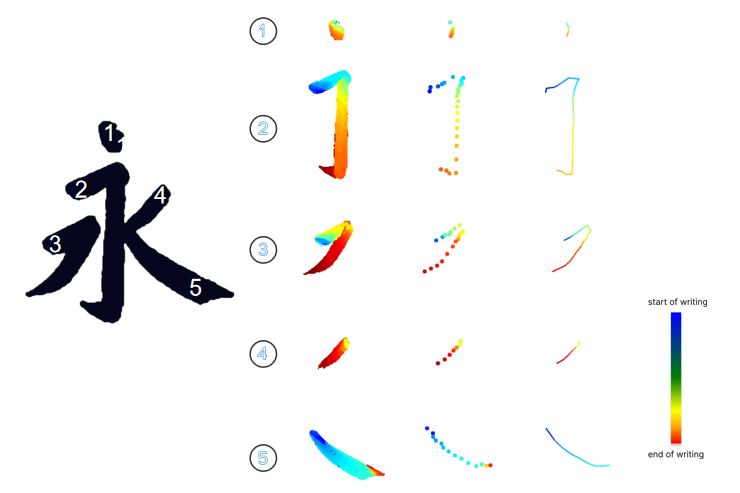}
    \caption{Skeleton extraction of each stroke in the character ``Yong''}
    \label{fig: stroke2skeleton}
\end{figure}

Chinese calligraphy is inherently an art of lines. To clearly highlight key positions in visualizations for students, extracting the skeleton of Chinese characters became a key focus of our work. 
We apply a centroid-based algorithm on the incremental ink deposition to extract the skeleton (Figure \ref{fig: Ink increment}). By connecting centroids over time, a smooth skeleton structure is obtained (Figure \ref{fig: stroke2skeleton}). Ultimately, the writing speed is represented by the offset of the ink's center point in each time frame.

\section{Visualization Design}
\begin{figure}[t]
    \centering
    \includegraphics[width=\linewidth]{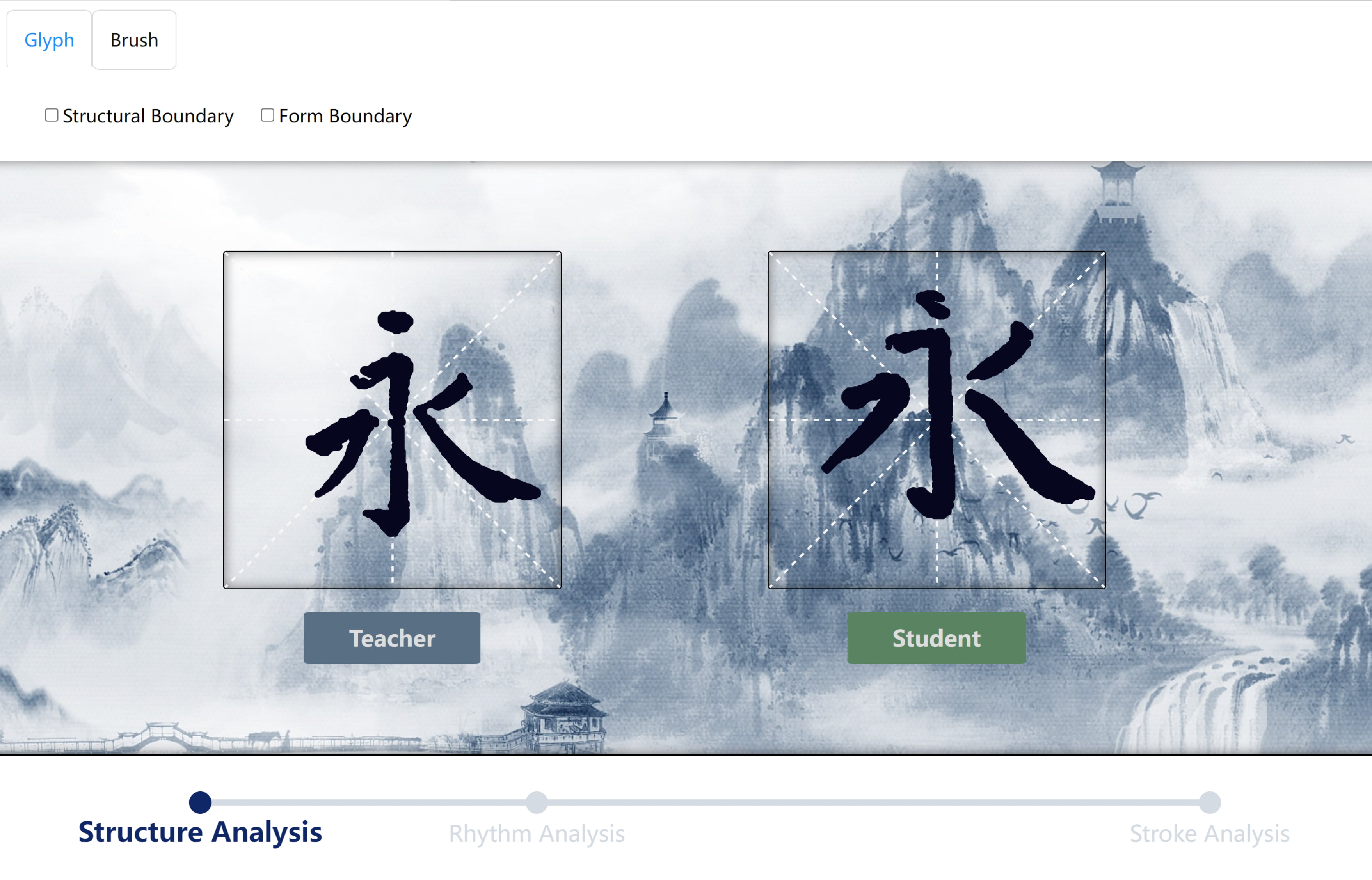}
    \caption{Step 1: Character Shape Comparison. The left side shows the teacher's writing, while the right side displays the student's work. Students can use the teacher's writing as a reference to identify structural issues in their own writing. Additionally, they can choose to use two types of guidelines—``Structural Boundary'' and ``Form Boundary''—to assist in their assessment.}
    \label{fig:Structure Analysis}
\end{figure}

Our visualization framework sequentially covers glyph observation, rhythm observation, and detailed stroke analysis. During the iterative design process, expert feedback was incorporated, emphasizing that calligraphy instruction typically begins with an overall view of the character's form.
 Therefore, the initial interface displays only the character's overall structure, concealing brushwork details (Figure \ref{fig:Structure Analysis}). This design encourages learners to first understand the overall composition and identify gaps between their work and the teacher's, allowing for targeted improvement in the next stage. This approach aligns with the cognitive process of traditional Chinese calligraphy training~\cite{dong2008creation}. Starting with glyph critiques also helps beginners unfamiliar with brushwork techniques to build confidence in a domain they may find challenging.

\begin{figure}[htbp]
    \centering
    \includegraphics[width=\linewidth]{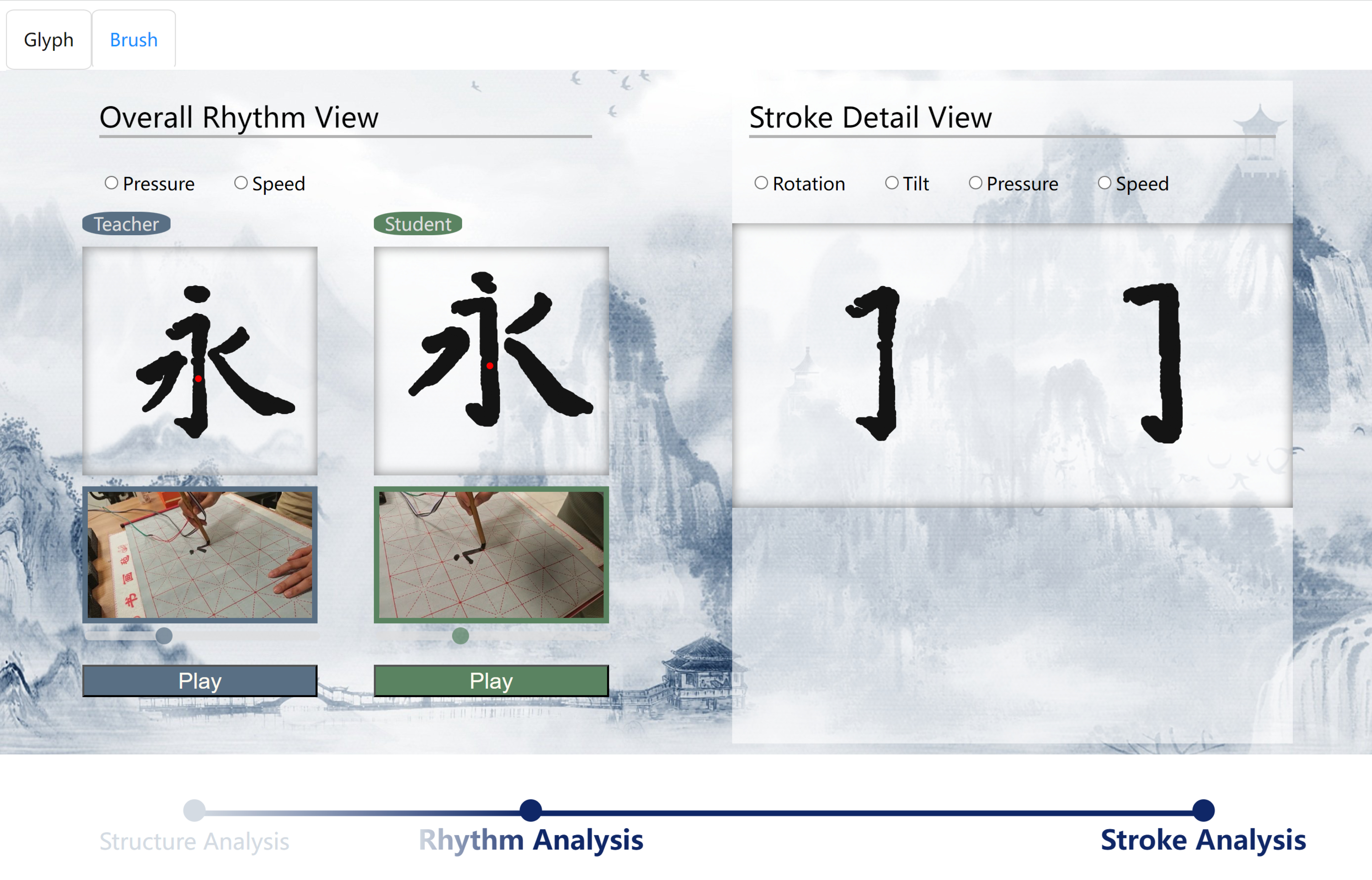}
    \caption{Step 2 and Step 3: Rhythm Analysis(Left) and Stroke Analysis (Right). These steps allow for the analysis of the speed and force variations throughout the writing of a complete character on the left, and the detailed brush technique used for a specific stroke on the right.}
    \label{fig:Rhythm and Stroke Analysis}
\end{figure}

During the iterative process, an interesting phenomenon was discovered: rhythm changes in dance and music were frequently used by calligraphy experts to elucidate the importance of varying writing pressure and speed. Calligraphy and dance are often compared in studies due to their similar rhythmic qualities~\cite{szeto2010calligraphic}. Building on the conclusions from the previous formative study\textbf{ (DC3)}, an overview of writing rhythm is introduced after the character structure analysis. Users can then select specific strokes and proceed to the next interface to analyze individual lines (Figure \ref{fig:Rhythm and Stroke Analysis}). This design also follows the standard framework of information visualization: (a) provide an overview first, (b) allow zoom and filter, and (c) offer details on demand\cite{shneiderman2003eyes}. Overall, a comparison between each part for the teacher and the student was provided, enabling the student to identify where the issues with posture are present \textbf{(DC5)}.

\subsection{Glyph Comparison}

To provide a more intuitive comparison between the teacher and student, we added the commonly used guiding grid – the ``Mi Zi Ge'' grid\cite{chinese_calligraphy_grids}. The grid not only serves as a reference but also helps observe the relative positions of the strokes.

\begin{figure}[H]
    \centering
    \includegraphics[width=\linewidth]{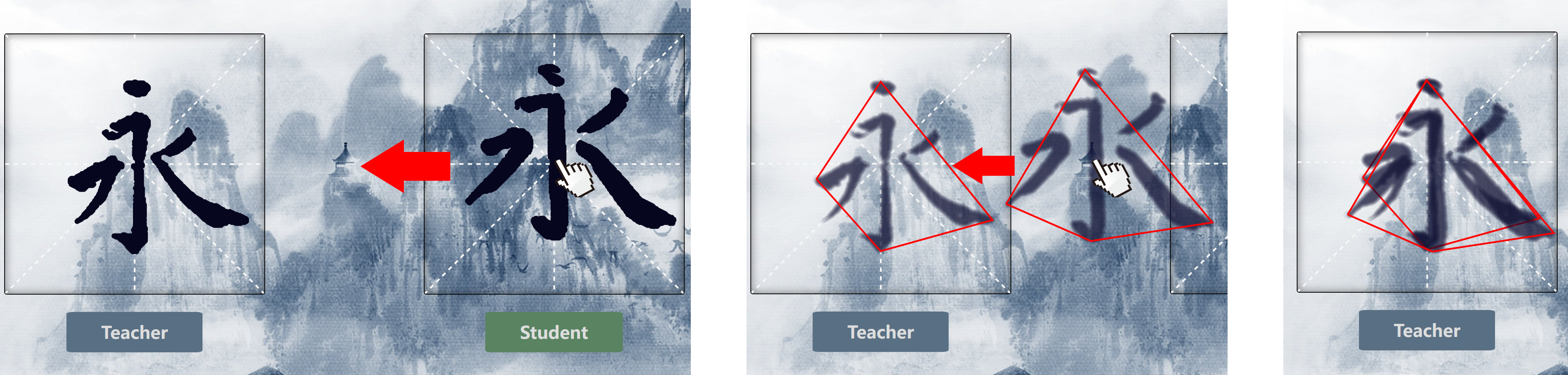}
    \caption{Compare the structure of students' and teachers' handwriting through dragging.}
    \label{fig:drag}
\end{figure}

\begin{figure}[H]
    \centering

    \includegraphics[width=\linewidth]{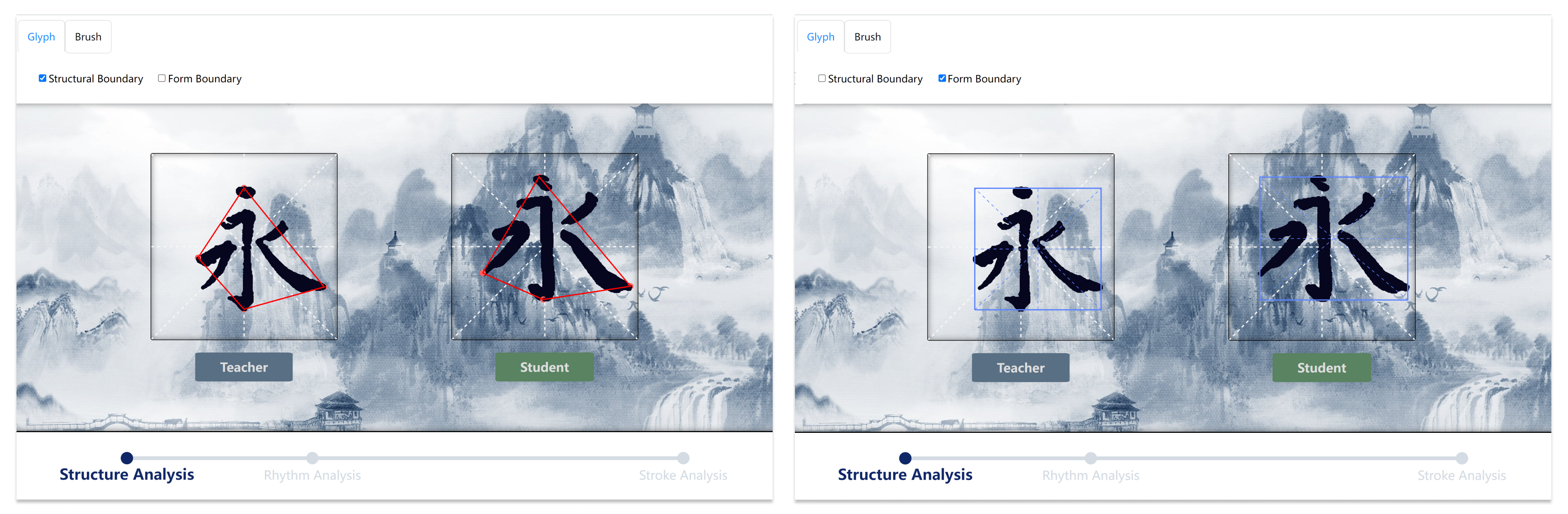}
    \caption{Two types of bounding boxes observe the structure of Chinese characters from different perspectives.}
    \label{fig:bounding box}
\end{figure}

Additionally, two optional boundary boxes were incorporated (Figure \ref{fig:bounding box}). The first boundary connects the four extremities of the character, showing the relative positions of the strokes. The second boundary uses these four points as edges to form a grid, allowing for a clearer view of the character's proportions and its central position. With these tools, the teacher and student can drag and compare characters to further analyze differences in writing techniques (Figure \ref{fig:drag}).

\subsection{Rhythm Comparison}

\begin{figure}[H]
    \centering
    \includegraphics[width=0.5\textwidth]{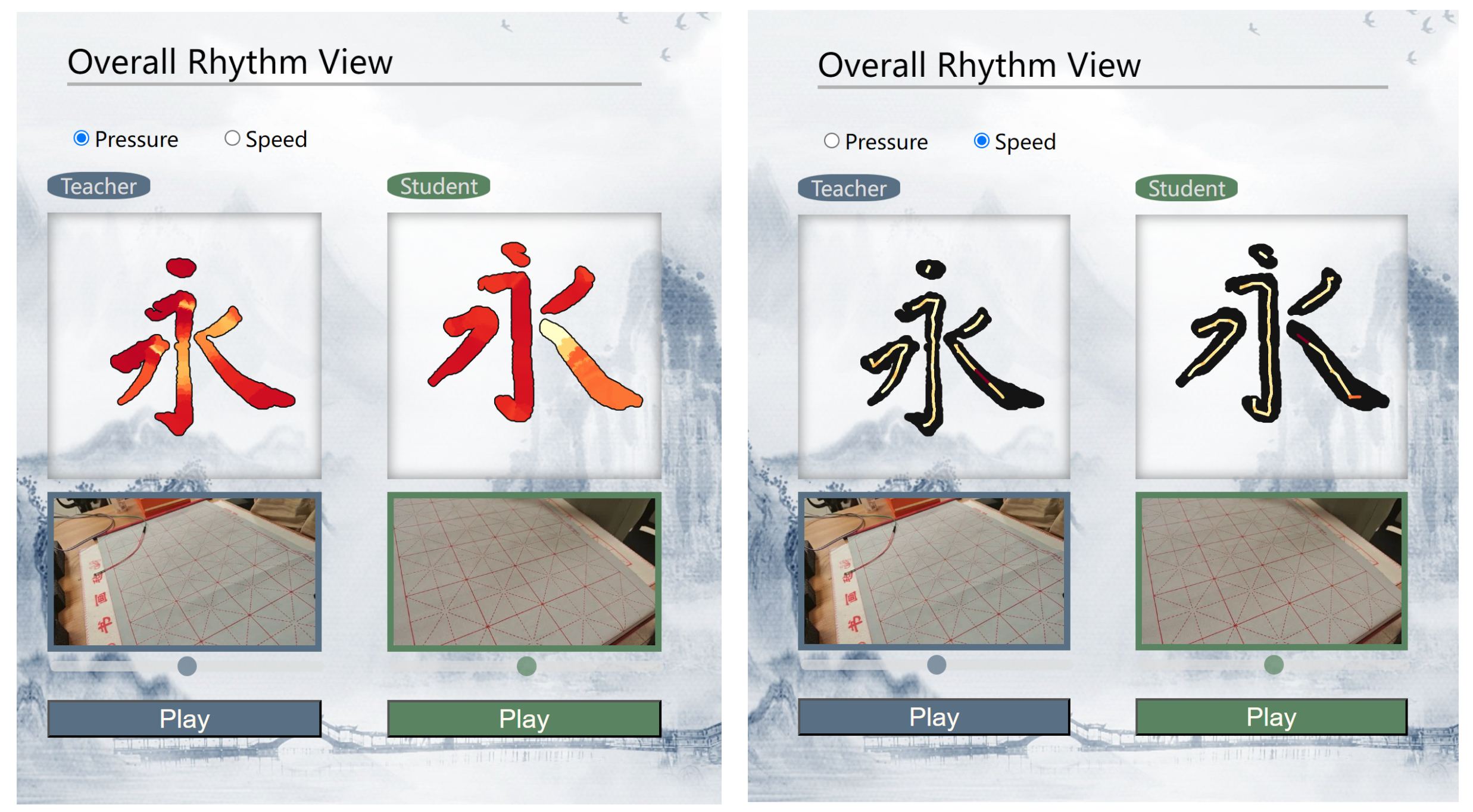}
    \caption{In the rhythm view, observe the changes in the teacher's and student's finger pressure (left) and speed (right) throughout the entire character.}
    \label{fig: Rhythm}
\end{figure}

To minimize confusion, finger pressure is represented using a heatmap, while speed is color-coded on the skeleton (Figure \ref{fig: Rhythm}) ~\cite{howe1983temporal}. In Figure \ref{fig: Rhythm}, the teacher's finger pressure varies throughout the stroke, typically applying force at the start and turning points. In contrast, the student's grip remained tense throughout the earlier strokes, only showing improvement in the final stroke. Similarly, speed can be analyzed in this way.

\begin{figure}[H]
    \centering
    \includegraphics[width=0.3\textwidth]{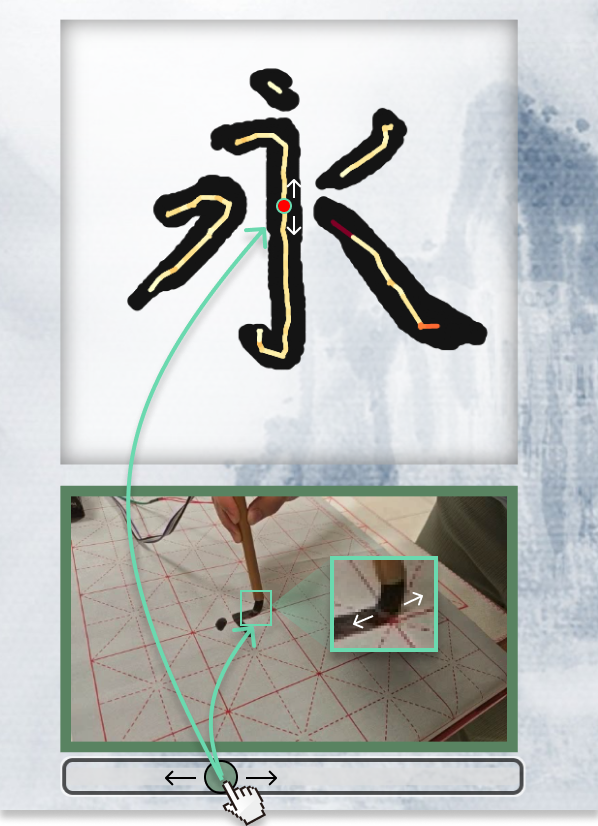}
    \caption{Drag the timeline to view the brush tip status at specific points of the character.}
    \label{fig:time bar}
\end{figure}

Center stroke (zhongfeng) is a fundamental principle in Chinese calligraphy ~\cite{yi2021beginner, yang2009animating, yang2013animating}, and therefore requires direct observation. To facilitate this, the writing video was retained, allowing teachers to integrate the brushstroke process with the form of the center stroke technique (Figure \ref{fig:time bar}). This helps students understand and connect the two aspects more effectively.

\subsection{Line Quality Analysis}

Placed on the same page as the previous rhythm analysis view, allowing users to observe brushstroke parameters while simultaneously comparing changes in the brush-tip video. This integrated approach enables a comprehensive analysis of stroke quality and the brushwork process.

\subsubsection{Brush Rotation}
In Chinese calligraphy, to maintain the ``center stroke'' (zhongfeng) and ensure smooth writing, the calligrapher must continuously adjust the position of the brush tip on the paper\cite{chiang1974chinese}. This adjustment is achieved by rotating the brush handle: on the one hand, the rotation ensures that the brush tip remains aligned with the center of the stroke, and on the other hand, when the brush tip begins to splay, the rotation consolidates the tip, ensuring that the force is concentrated at the tip. This focus on brush handle rotation reflects the calligrapher's exploration of brush-tip control. Therefore, the rotation of the brush handle in writing is worth demonstrating.

Initially, an attempt was made to decompose the posture of the brush handle into yaw, roll, and pitch\cite{fitzpatrick2010validation}, which is a common method for analyzing movement. The idea of displaying these values on a dashboard was considered, but isolated parameters at a single moment in time provided little explanatory value. We also explored displaying the brush handle posture curves (e.g., yaw, roll, and pitch) for both teachers and students side by side \cite{10.1145/3476124.3488645, 10.1145/1878083.1878098}. However, it was found that these curves overly abstracted the posture information, rendering them less accessible to users without a background in data visualization.

\begin{figure}[htbp]
    \centering
    \includegraphics[width=0.8\linewidth]{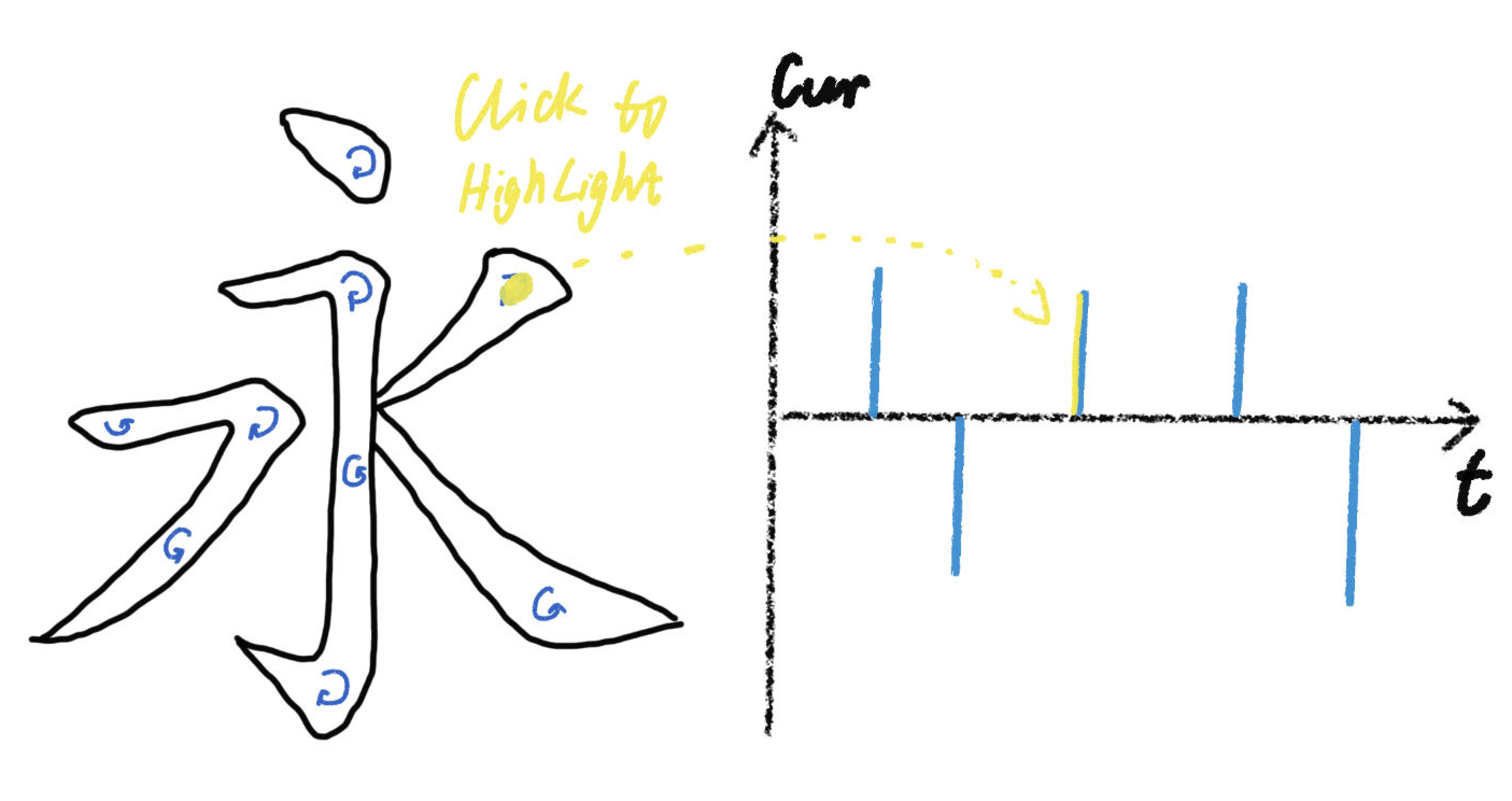}
    \caption{One of the sketches during the iterative process: A rotating arrow is used to indicate that a rotation has occurred at a certain position in the handwriting, and this is represented with a pulse diagram.}
    \label{fig:roll}
\end{figure}

Ultimately, a decision was made to visualize the rotation directly on the written strokes. Initially, the plan was to capture and display the amount of rotation at corresponding locations (Figure \ref{fig:roll}). However, given the possibility of brush rotation during advancement, marking the absolute orientation of the brush handle at sampled points along the stroke's central axis using arrow symbols was chosen. This approach offers a more intuitive understanding of the rotation process.

\begin{figure}[htbp]
    \centering
    \includegraphics[width=0.8\linewidth]{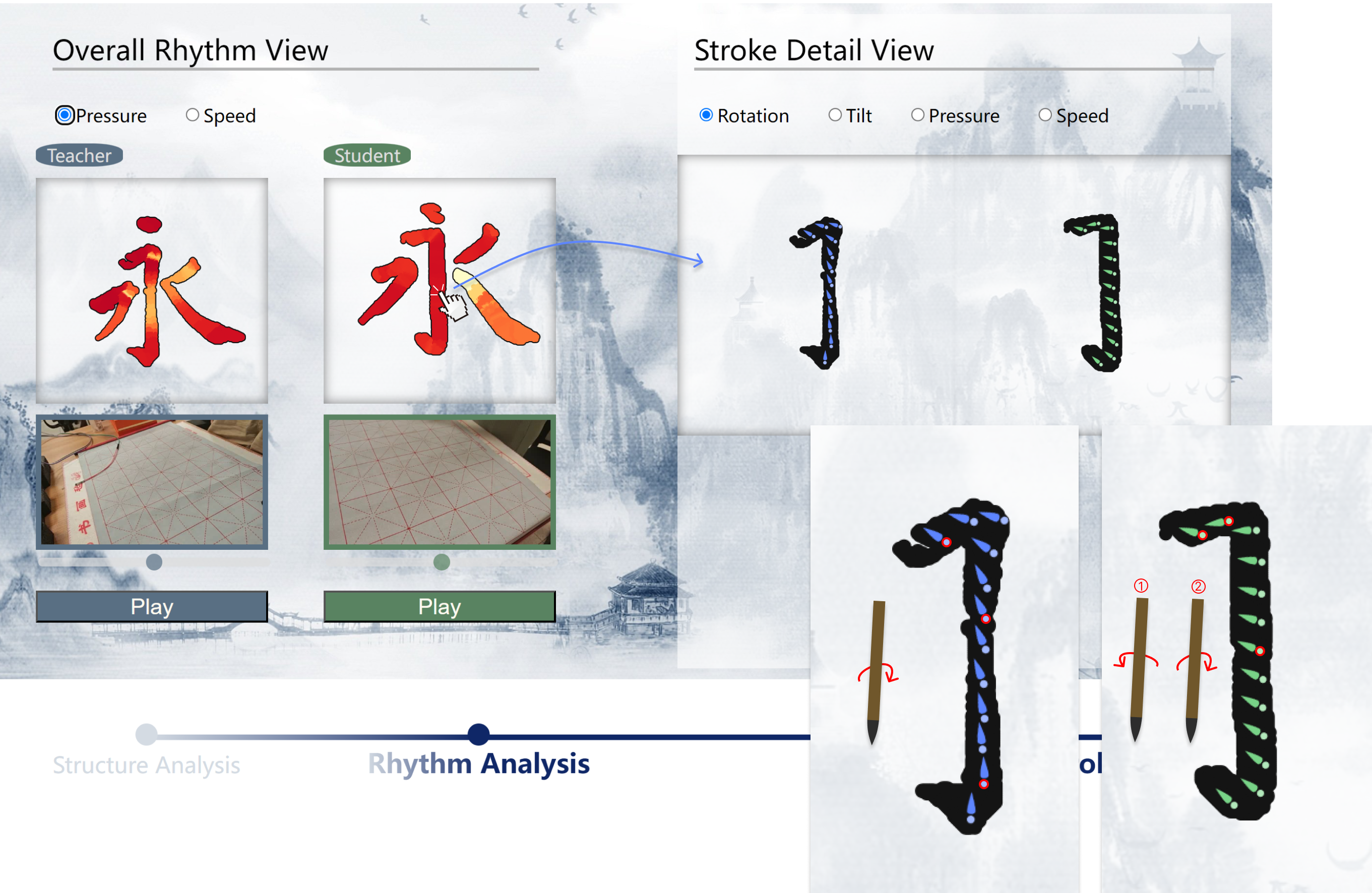}
    \caption{In the stroke detail view, the brush rotation of both the student and the teacher while writing the same stroke is visualized through the rotation view.}
    \label{fig:Rotation}
\end{figure}

Given the cylindrical nature of the brush shaft and the absence of a fixed front-facing direction, the reverse direction of the first stroke's extension is adopted as the initial orientation, aligning with customary writing practices. The calibration is done at the start of each stroke (Figure \ref{fig:Rotation}). Figure \ref{fig:Rotation} shows that the teacher rotates (left) the brush clockwise while advancing it, whereas the student (rignt) first rotates the brush counterclockwise, followed by a slight clockwise rotation, without achieving the same 'brush wrapping' technique as the teacher.

\subsubsection{Brush Tilt}
\begin{figure}[ht]
        \centering
        \includegraphics[width=0.8\linewidth]{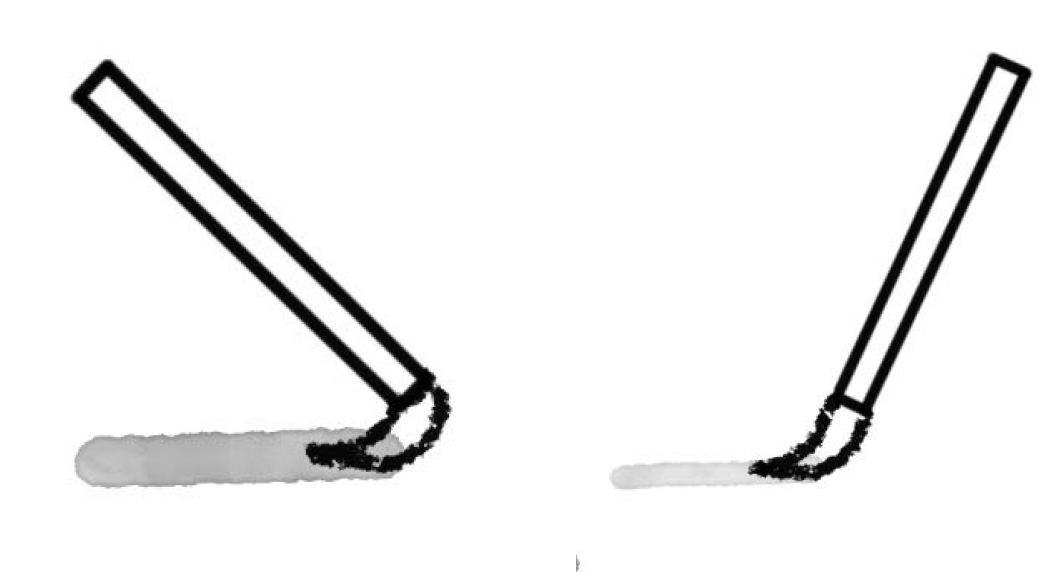}
        \caption{The texture of lines varies with two different pen angles.}
        \label{fig:brush tilt}
\end{figure}

\begin{figure}[th]
        \centering
        \includegraphics[width=0.98\linewidth]{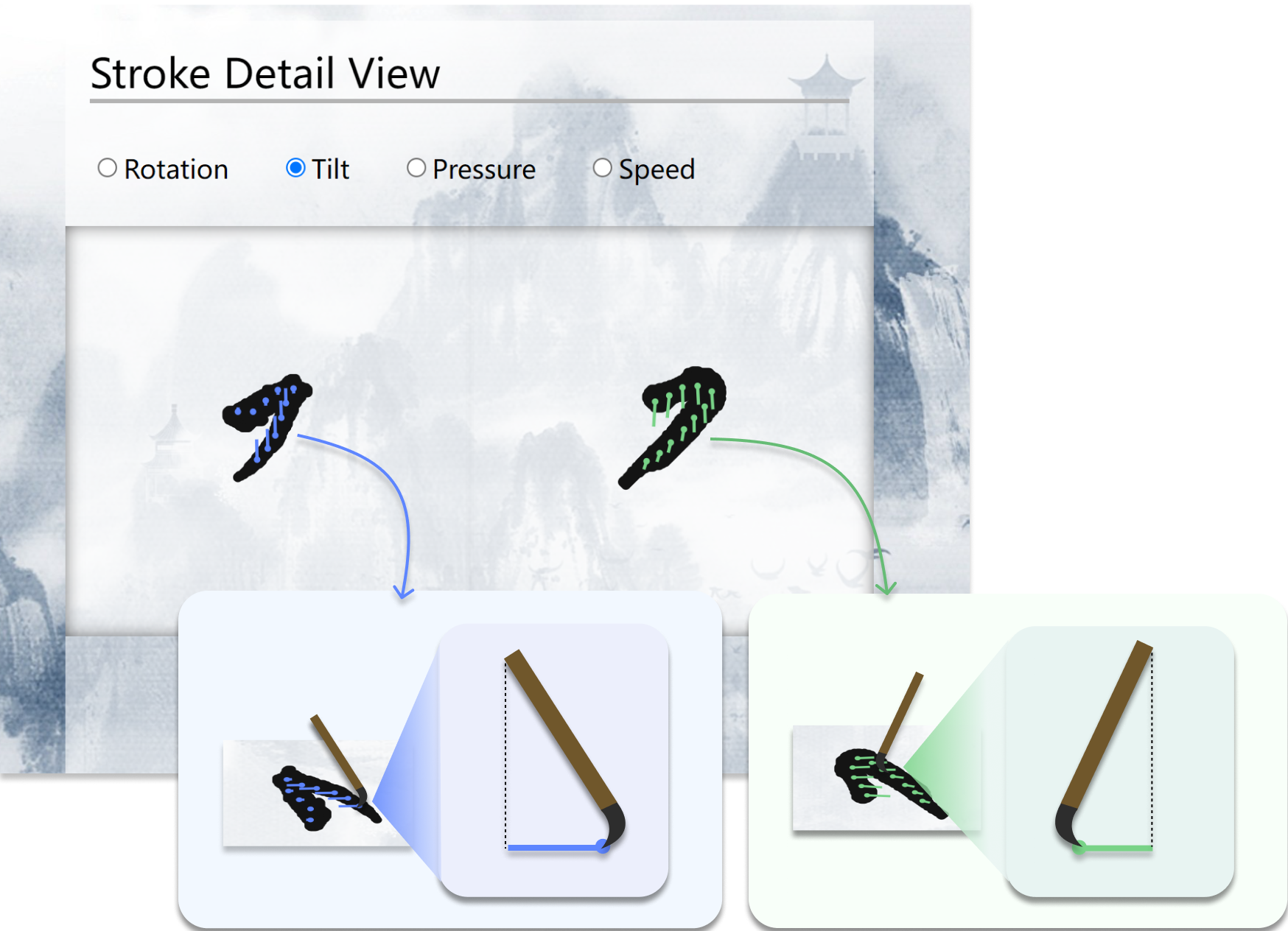}
        \caption{The short lines on the strokes represent the brush's projection on the paper. From the visualization, it's clear that the brush's tilt direction differs between the teacher and the student when writing the same stroke. The teacher tends to tilt the brush against the writing direction to add more strength to the lines, while the student completely overlooks this technique.}
        \label{fig:stroke splitting}
\end{figure}

The tilt angle of the brush handle directly affects the way the brush tip interacts with the paper (Figure \ref{fig:brush tilt}), thereby altering the friction during writing, which significantly impacts the quality of the strokes. Typically, strokes with greater friction appear darker, thicker, and have sharper edges, while those with less friction are lighter, thinner, and have more blurry edges (Figure \ref{fig:stroke splitting}). Therefore, to explore the conditions that contribute to the texture of a particular stroke, it is essential to display the tilt of the brush handle in various directions. Building on the discussion from the previous section, the tilt direction of the brush handle must be directly visualized on the stroke.

To ensure ease of comprehension for users, a 3D model of the brush hovering above each sampled point on the stroke was initially planned to be rendered to represent the brush's position at that location. However, since the strokes lie on a horizontal plane, the 3D brush could become obscured or cause perspective distortion due to overlapping brushes in the foreground and background \cite{munzner2015visualization}. As a result, the projection of the brush handle onto the plane of the paper was ultimately chosen as the most suitable visualization method.

\subsubsection{Comparison Curves for Speed and Pressure}
The speed of the brush and the pressure applied by the fingers are critical variables that influence the quality of calligraphy strokes. These two factors respectively determine the force of the brush and the duration of contact with the paper surface. The visualization of pressure and speed parameters in this view continued with the encoding approach from the rhythm view. The difference is that the upper and lower limits of the color scale were adjusted for each stroke to enhance the visibility of pressure variations within individual strokes. Considering that both pressure and speed are unidimensional numerical variables, they are better suited for comparison through curves rather than posture data. To prevent confusion and decrease cognitive load~\cite{keller2006information}, scatter plots were selected to illustrate pressure changes, while curves were used to represent speed trends.

Calligraphy does not follow rigid brushwork rules; much like riding a bicycle, where direction is adjusted based on road conditions, the writing process requires continuous adjustments in brush posture based on the state of the brush. Thus, the primary requirement was to display the general range and trends. In the charts, specific numerical values for pressure and speed were omitted, and a tiered representation was chosen instead.

First, the pressure view is introduced, wherein the bottom scatter plot displays the range of pressure levels for both teachers and students. When hovering the mouse over a scatter point, a small red dot will appear on the stroke to indicate the corresponding position on the ink trace. The x-axis of the scatter plot represents the position within the stroke, and scatter points aligned vertically represent roughly the same position on the stroke, facilitating comparison. As the mouse moves across the chart, a vertical line and the corresponding red dot on the stroke will move in sync, indicating the matching position on the ink trace.

\begin{figure}[t]
    \centering
    \includegraphics[width=\linewidth]{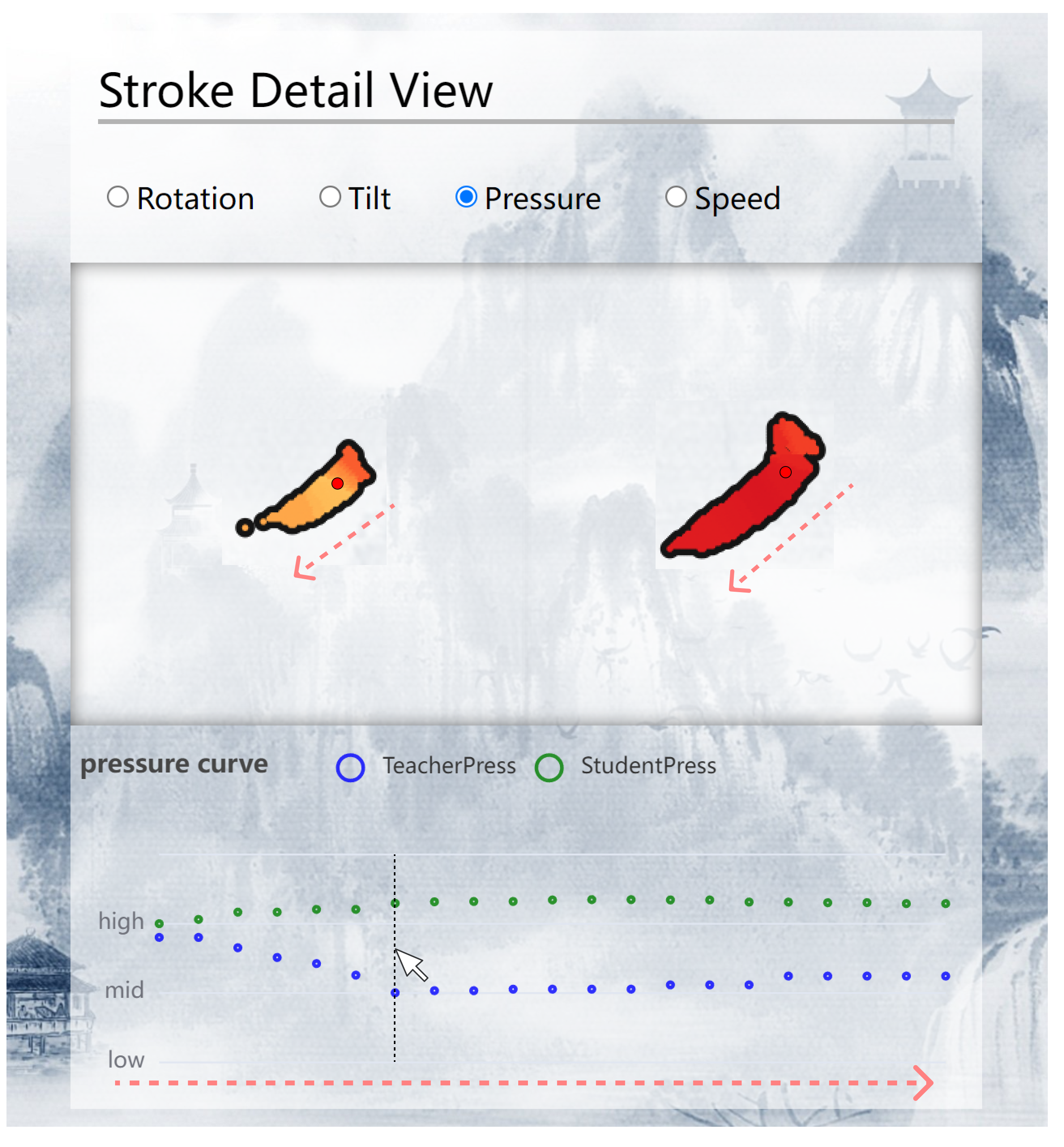}
    \caption{electing a left-falling stroke for pressure analysis, the scatter plot clearly shows the pressure variation trends and the differences between the student and the teacher. The X-axis represents the relative position of the stroke, and users can hover over the vertical line of any sample point. Simultaneously, a small red dot in the character will be positioned at the corresponding location, making comparison and analysis easier.}
    \label{fig: Pressure}
\end{figure}

Through the visualization, we can observe that when writing a short downward stroke (pie), the teacher applies greater pressure with the fingers in the initial phase, gradually relaxing to maintain a moderate pressure level (Figure \ref{fig: Pressure}). In contrast, the student consistently applies high pressure throughout the stroke. The hover function was utilized to identify the point of maximum pressure difference between the teacher and the student, which was approximately located at the 2/5 mark of the stroke. This feature provides clear guidance on the area where the student requires adjustment.

\begin{figure}[H]
    \centering
    \includegraphics[width=\linewidth]{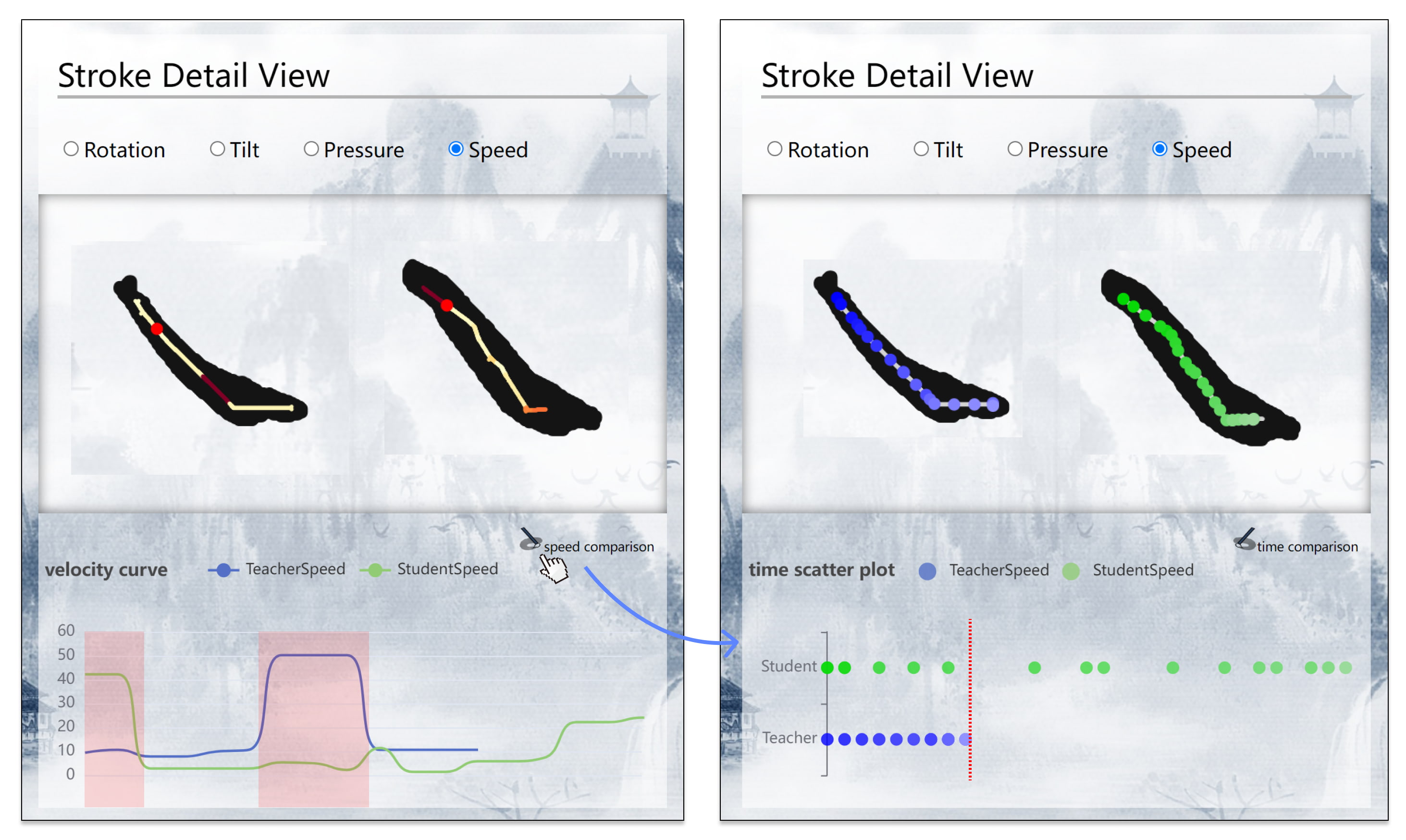}
    \caption{By comparing the writing speed of the teacher and the student in the same stroke segment using line charts (left) and their writing positions at the same time points (right). }
    \label{ref: Speed}
\end{figure}

Next is the speed view, where the X-axis also represents the stroke position. The curve shows the speed variation throughout the stroke, allowing for a comparison with the teacher's speed (Figure \ref{ref: Speed}). The aim was also to observe the variation in writing speeds of the teacher and the student over time. To accomplish this, time was set as the x-axis, analogous to a stopwatch, with two rows of horizontal scatter points representing the writing progress of both the teacher and the student. Furthermore, color saturation was employed as an additional layer of encoding to align the sequence between the top and bottom of the ink trace and the graph.

In the speed comparison (left), the teacher's writing shows a rhythm of faster movement in the middle and slower at both ends, while the student writes quickly at the start, with little variation in speed afterward. This indicates that the student's initial and finishing strokes are not executed properly. However, the right chart reveals that the student's overall writing speed is three times slower than the teacher's, suggesting hesitation at the beginning, which leads to sluggish and lifeless strokes \cite{chiang1974chinese}.

\section{User Study}
The user study analyzes our system from two perspectives. First, the study will assess whether students can enhance their understanding of basic brush techniques through visualization and subsequently improve their calligraphy skills. Second, it will examine whether the tool effectively aids teachers in conveying complex calligraphy movements to students, thereby improving teaching efficiency. We compare the results of teaching the same skill both with and without the assistance of the CalliSense system, using the non-CalliSense method as the baseline.

\subsection{Participants}
We recruited participants through electronic posters, flyers, and the snowball sampling method~\cite{naderifar2017snowball}. A total of 6 calligraphy teachers and 28 students with varying levels of calligraphy experience and professional backgrounds signed up. After screening, we selected 4 teachers (T1-T4) with 1 to 2 years of teaching experience, whose teaching abilities were comparable and sufficient to cover the most basic calligraphy concepts. Additionally, we chose 12 students (S1-S12) with similar backgrounds in calligraphy, all of whom had less than one year of practice experience. Each participant received a reward of 100 CNY (approximately 14.05 USD).

\subsection{Study Design and Procedure}
The 12 students were randomly assigned to the four teachers, with each teacher guiding three students. Four teaching workshops were conducted, where each teacher tried both the CalliSense-assisted and traditional teaching methods to complete two different teaching tasks. To assess the students' learning outcomes, we used a within-subjects design and administered questionnaires after each teaching condition to evaluate their understanding of basic calligraphy concepts. Following the experiment, semi-structured interviews were conducted with both the teachers and students.

When designing the user study process, we consulted a calligraphy expert with 35 years of writing and teaching experience and established two guiding principles:
1) Ensure that the teaching content represents widely agreed-upon basic calligraphy knowledge. As artistic creation can involve personal expression, we needed to confirm that our system is applicable to most teaching scenarios;
2) Ensure that the content is suitable for beginners to understand.

The expert also noted that while it is difficult to fully master brush techniques in just 30 minutes, it is sufficient to demonstrate what correct brushwork looks like. Therefore, our evaluation focused more on participants' conceptual understanding of brush techniques rather than their actual writing performance.

\begin{figure}[h]
    \centering
    \includegraphics[width=0.4\textwidth]{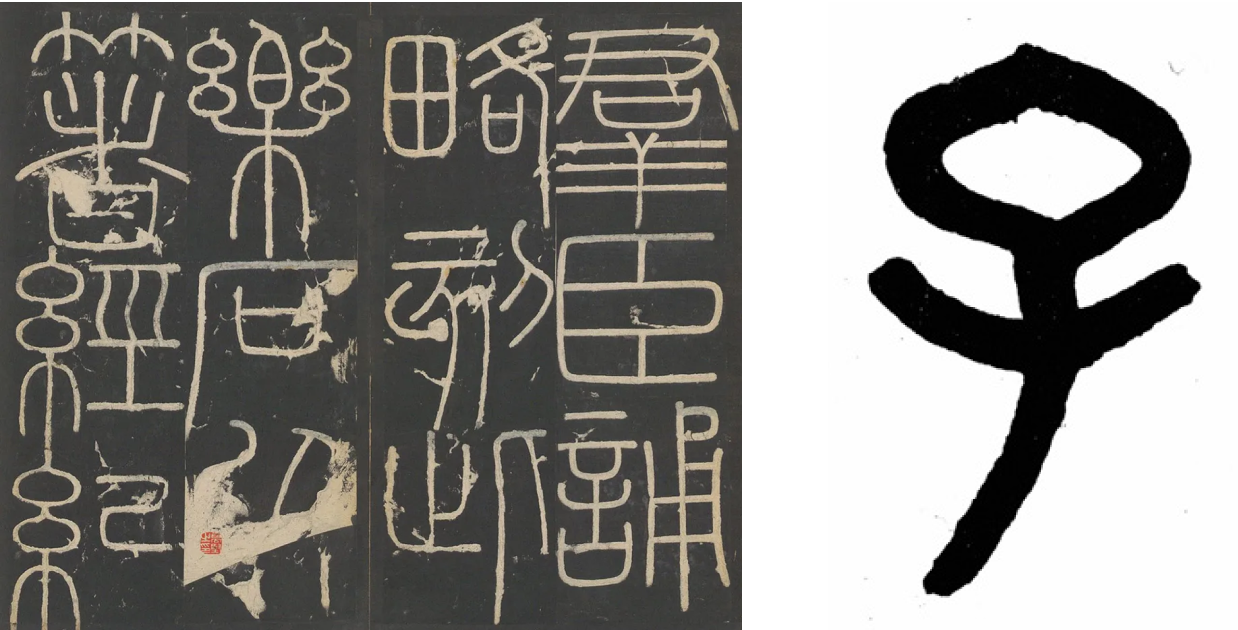}
    \caption{The Yishan Stele (left) and the selected character ``Zi'' from the Yishan Stele (right).}
    \label{fig:Yishan Stele}
\end{figure}

We designed the study to have equal-length sessions for the same teaching tasks to assess the system's effectiveness and efficiency. The teaching approach followed a common method where students write a complete Chinese character, and the teacher provides feedback and corrections. In our study, we specifically chose the ``Yishan Stele'' (Stele of Mountain Yi)~\cite{metmuseum_yishan_stele} as the practice subject. The ``Yishan Stele'', one of China's earliest stone inscriptions, is renowned for its uniform strokes and consistent brush techniques. Since it consists of simple strokes and uses relatively basic brush techniques, it is often used as introductory material for beginner calligraphers. Notably, the stele includes both curved and long strokes, offering an excellent opportunity to practice central-axis brushwork (zhongfeng). Given that our participants had limited calligraphy experience, we carefully selected the character ``Zi'' with a moderate number of strokes that incorporates both curved and long lines, allowing them to better grasp fundamental calligraphy skills (Figure \ref{fig:Yishan Stele}).

By practicing these strokes, learners can master two key concepts: 
1. The concept of ``rolling the brush'' (guofeng), where rotating the brush handle helps achieve the central-axis brushwork~\cite{wang2024formation}. 
2. Controlling the texture of the strokes by adjusting the writing speed and the tilt of the brush handle~\cite{qian2007towards, long2001art}.

\begin{figure}[t]
    \centering
    \includegraphics[width=0.44\textwidth]{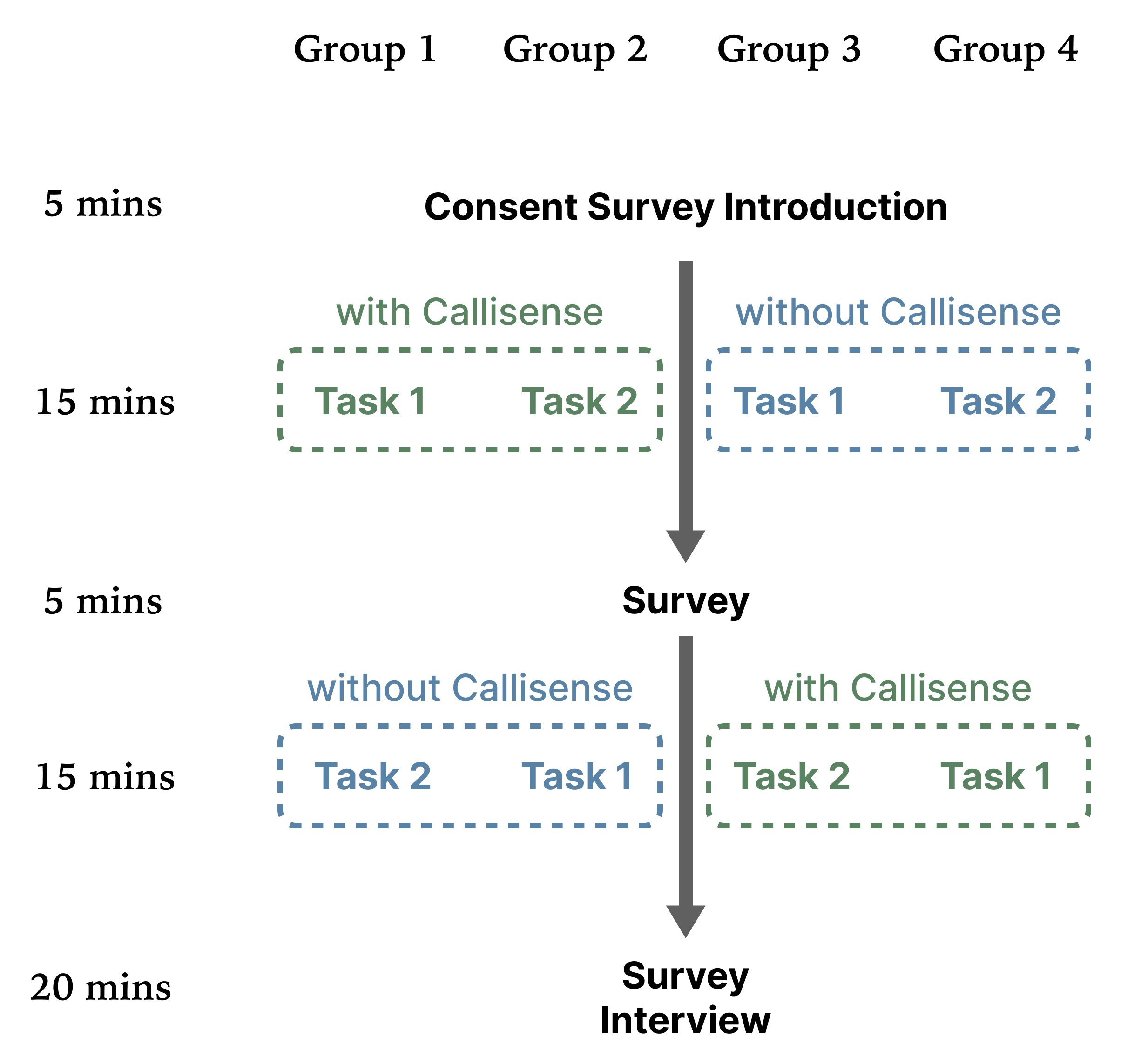}
    \caption{User Study Process. The four columns represent four different user research processes, which will be randomly assigned to four different teachers.}
    \label{fig:user study process}
\end{figure}

Based on this, we designed two teaching tasks aligned with common foundational calligraphy techniques. The first task focuses on teaching the ``central-axis'' (zhongfeng) brush technique, which involves the concept of ``rolling the brush'' (guofeng) by rotating the brush handle to concentrate the brush tip. The second task teaches the concept of ``controlling stroke texture'', where brush quality is managed through the tilt of the brush handle, writing speed, and hand pressure. While the brush technique parameters may not seem evenly distributed between the two tasks, the concept of ``rolling the brush'' is relatively difficult to grasp, as confirmed by pre-experiment test questionnaires. Therefore, the cognitive load required for both tasks is similar. To avoid learning effects, a Latin square design was used. The teaching tasks, combined with the presence or absence of the CalliSense system, created four distinct process flows (Figure \ref{fig:user study process}), which were randomly assigned to the four teachers.

The final procedure is as follows: Prior to the experiment, participants completed an informed consent form and a demographic survey. Students also took a pre-test to assess their knowledge of the calligraphy concepts covered in the teaching tasks, while teachers used this time to learn how to use the system. Afterward, the teachers conducted the first teaching task. Upon completion, students filled out a mid-task questionnaire to assess their understanding of the concepts taught in the first task, and they rated the class for engagement, involvement, and clarity. The second teaching task followed, and after it was completed, students filled out a post-task questionnaire to evaluate their understanding of the second task, along with class ratings.

Finally, both teachers and students completed the NASA TLX~\cite{hart2006nasa} and SUS questionnaires~\cite{10.5555/2835587.2835589} and participated in semi-structured interviews to discuss the differences between traditional teaching methods and CalliSense, their experience with the system, and how it specifically aided calligraphy learning. We also encouraged participants to think aloud during the experiment, and all verbal feedback was recorded for analysis.

\subsection{Analysis and Results}

\subsubsection{System Usage Overview:}

\begin{figure*}[h]
    \centering
        \includegraphics[width=0.49\textwidth]{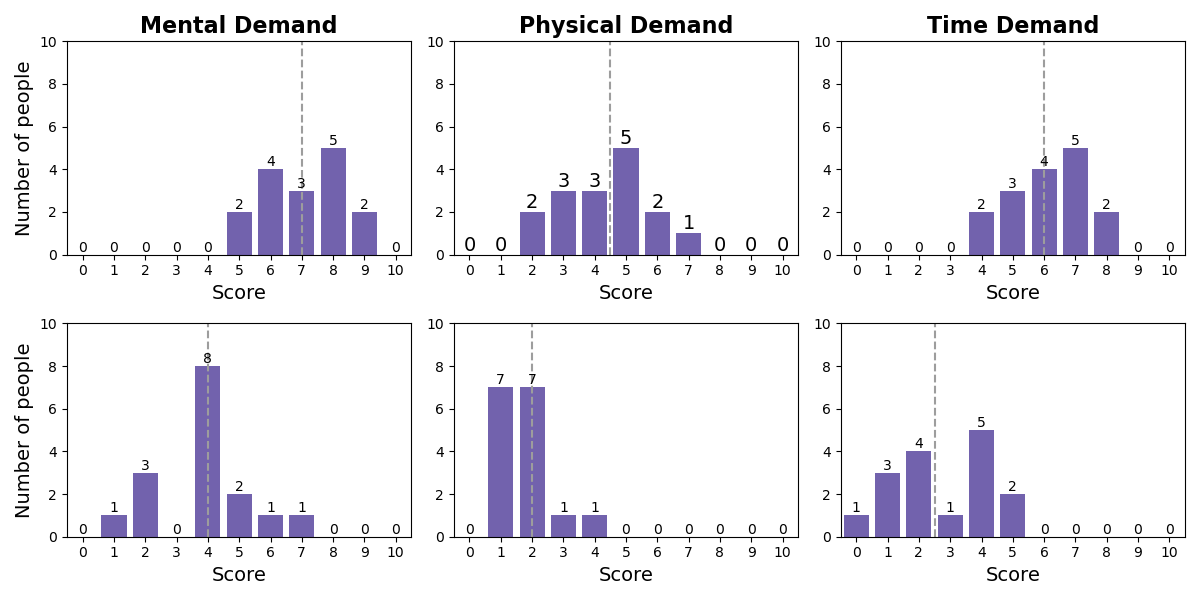}
        \includegraphics[width=0.49\textwidth]{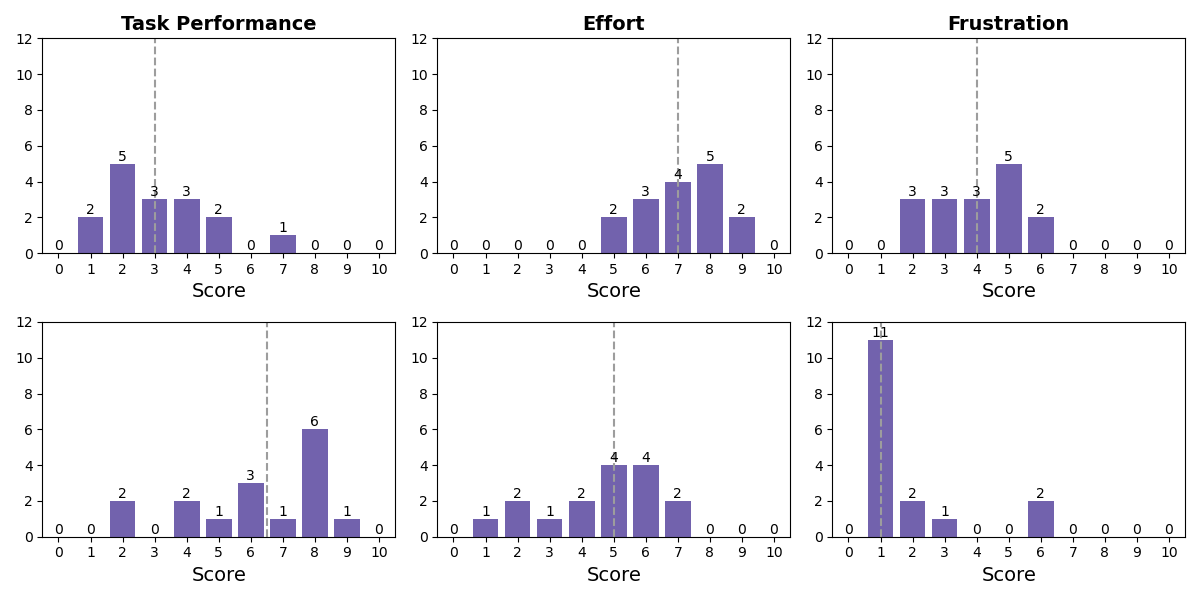}
    \caption{Participants rated system usability using the NASA Task Load Index (1-10 scale), Top: Ratings for the traditional classroom, Bottom: Ratings for the classroom using CalliSense.}
    \label{fig: NASA}
\end{figure*}

Overall, both teachers and students agreed that CalliSense effectively aids students in mastering Chinese calligraphy techniques. Using the NASA TLX questionnaire (scale 1–10), we evaluated the system's impact on teaching effectiveness (Figure \ref{fig: NASA}). In terms of performance (brushwork learning), students showed significant improvement in their brush technique when using CalliSense (Md=8, IQR=3) compared to the baseline (without CalliSense) (Md=4, IQR=2). They were able to pinpoint specific areas for improvement instead of relying on vague impressions. One participant even brought their own brush and paper, attempting to recreate their usual practice setup with the system's sensors. They described the experience as viewing their practice from a fresh perspective (S5).

Frustration levels were generally low (Md=1, IQR=1), although one student remarked that, despite initially thinking their brushwork was quite good, the experience revealed overlooked issues. This student was surprised by the number of brushwork details that needed attention but saw this as a positive discovery in their calligraphy learning (S9). Both mental (Md=4, IQR=2.5) and physical demands (Md=2, IQR=1) were notably reduced. Another student mentioned that the ability to clearly identify areas for improvement reduced their previous anxiety (S10). Participants generally found CalliSense enjoyable to use, with improvements in time demands (Md=2.5, IQR=3) and effort (Md=5, IQR=4). However, effort varied significantly, mainly due to the complexity of the ``wrapped brush'' technique, which created individual differences based on experience and learning strategies (S10).

Additionally, the SUS questionnaire was used to evaluate system usability, with an average score of 78, significantly higher than the benchmark of 68, indicating good usability. Apart from the interface's ability to convey information clearly, the sensors were considered lightweight and barely noticeable, especially on the fingers (N=6). However, one teacher noted that the wire at the top of the inertial sensor could interfere with cursive script writing, which requires more fluid brush movements (T4).

\subsubsection{Knowledge Mastery Evaluation:}

\begin{figure}[t]
    \centering
    \includegraphics[width=0.4\textwidth]{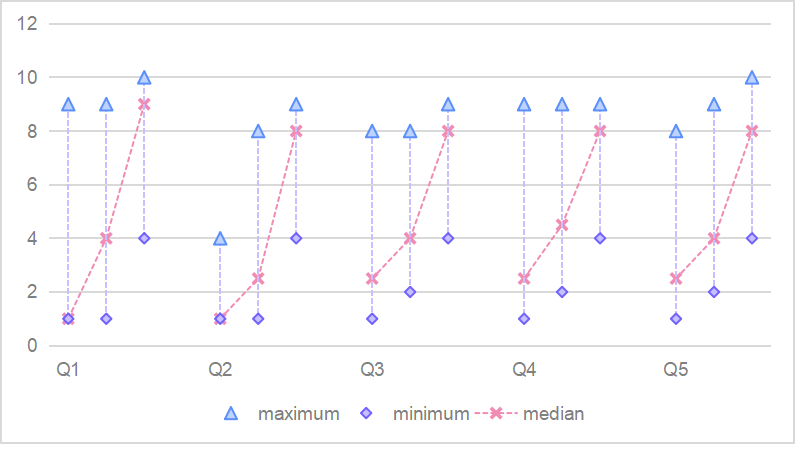}
    \caption{User research results show that students' mastery of 5 calligraphy brush knowledge points after learning in two different classroom settings. Compared to traditional classrooms, students' learning outcomes for the 5 knowledge points have significantly improved in classrooms equipped with CalliSense.}
    \label{fig:knowledge points test}
\end{figure}

We also examined how well students grasped the key concepts in the instructional tasks, asking them to self-assess their understanding of five fundamental brushwork techniques: Q1 - using the brush's center stroke (zhongfeng), Q2 - the ``encircling'' technique (guofeng), Q3 - how brush tilt affects line quality, Q4 - the impact of hand pressure on line formation, and Q5 - how writing speed influences the stroke. These aspects encompass all the parameters demonstrated by the CalliSense system. The results showed significant improvement in these five dimensions when CalliSense was used, compared to traditional classroom instruction (Figure \ref{fig:knowledge points test}). This improvement can be attributed to the system's ability to visually present data, allowing students to clearly identify areas for improvement and track their progress.

Some students even noted that with the recorded data from the system, they no longer needed constant guidance from the teacher and could independently focus on targeted practice, such as writing speed and brush angle (S7, S12). Additionally, several students mentioned that just one use of the system was enough to shift their understanding, helping them recognize which aspects were most critical and where they should focus more attention. For instance, some students had not previously understood the importance of writing speed, but the system-generated charts made this concept clear to them (S3). The structured visualization not only summarized the entire practice process from beginner to advanced levels but also highlighted that enhancing awareness of brush control details was more crucial than simply practicing individual strokes~\cite{zhang2023bringing}.

Furthermore, students appreciated the comparison feature in CalliSense, which allowed them to consciously replicate their teacher's movements (S6). Interestingly, one student independently discovered correlations between different visualized parameters, such as noticing that hand pressure decreases during brush rotation because the ``encircling'' technique requires a more relaxed grip (S11). This realization deepened their understanding of how grip tension affects the quality of the strokes. These findings suggest that CalliSense not only helps students master specific skills but also encourages them to think critically about brush control, thereby enhancing their overall understanding of calligraphy techniques.

\subsubsection{Classroom Experience:}

Students' subjective experience in class can influence how effectively they absorb knowledge. We asked students to rate two classroom experiences on scales including ``Q6: I found the class enjoyable'', ``Q7: I found the class relaxing'' and ``Q8: I felt engaged in the learning process'' with scores ranging from 1 (strongly disagree) to 10 (strongly agree). Results showed that classes with CalliSense scored higher in Q6-enjoyment (Md=8, IQR=2) and Q8-engagement (Md=7, IQR=2.5) compared to baseline classes. In class, students were eager to use the writing equipment, curious to see how their brushwork would be visualized. The immediate feedback from the system significantly boosted their engagement. After each writing session, they were able to see their process right away, and the concrete feedback from visualizations was more persuasive than vague praise or self-assessment (S1, S11). While Q7-relaxation (Md=7, IQR=3) was similar between CalliSense and baseline classes, both scored relatively high.

\subsubsection{Teacher Feedback:}

Teacher insights were mainly gathered through interviews. Teachers generally agreed that without tools like CalliSense, it was difficult to observe so many fine details (N=4). The system's visualizations helped them better understand students' writing processes. However, one teacher noted that using the system might require adjusting their teaching strategy (T2). They usually help beginners by repeatedly demonstrating ``incomplete strokes'' rather than analyzing each stroke individually. With the introduction of the system, they may need to rethink this approach. Still, they believed the adjustment was worthwhile, as the system allowed them to focus more on the core elements of calligraphy learning.

Some teachers also admitted that they had not previously paid enough attention to students' pressure control when holding the brush but now realized its critical importance (N=2). Another teacher highly appreciated CalliSense, seeing its potential for online teaching (T3). Moreover, the system's strengths in personalized teaching became evident. As each student faced unique challenges in writing, the system's detailed brushwork parameters allowed teachers to pinpoint specific weaknesses, rather than providing broad guidance. This targeted instruction not only improved teaching efficiency but also accelerated students' progress in calligraphy.

\section{Discussion}

\subsection{Capturing Brush Motion in Calligraphy}
When discussing techniques for capturing the process of brush calligraphy, we evaluated multiple possibilities and ultimately selected a solution that meets research needs while balancing flexibility and cost-effectiveness.

Firstly, we observed that hand movements exhibit minimal explicit variation (as the hand maintains a consistent grip on the brush), so capturing hand gestures was not prioritized. Instead, the movement characteristics of the brush itself stood out as the primary focus of our research. The brush shaft, with its relatively simple shape, is easier to abstract and model, prompting us to explore various techniques for capturing its motion.

One direct approach involves inferring the brush’s motion trajectory by analyzing the written output. For example, key positions such as stroke starting and ending points can be processed systematically to reconstruct writing movements~\cite{10.1145/3526114.3558657, 10.1145/3613904.3642792}. While this method is effective for creating stylistically consistent fonts, it falls short in capturing the personalized nuances in learners’ writing and the subtle motions of the brush. Another approach is leveraging optical motion capture devices, such as Leap Motion~\cite{weichert2013analysis} to track the brush. By marking key points on the brush, these devices can measure its position and tilt~\cite{Matsumaru_2017jaciii, 10.1145/3029798.3038422}. This method offers exceptional precision (up to 0.01mm) but cannot directly capture brush rotation, requiring additional sensors for complete data. To minimize the complexity and cost of the equipment while maintaining a natural writing experience, we opted to reduce auxiliary device usage~\cite{chang2007simplicity, mcinnerney2004online}. However, in non-teaching scenarios focused on detailed analysis, this high-precision technology could still be considered.

Contact-based motion sensing devices, such as the PHANToM Omni~\cite{silva2009phantom} and Force Dimension Omega~\cite{chen2018patient}, provide an alternative by recording and reproducing writing motions while offering haptic feedback to help learners perceive subtle actions~\cite{10.1145/1255047.1255063, nishino2011calligraphy}. However, these devices are not specifically designed for brush calligraphy, often requiring additional force to operate, which may compromise the natural writing experience. Moreover, their high cost limits accessibility and widespread adoption.

After weighing the options, we ultimately chose to attach small inertial sensors to the brush. This approach captures critical spatial characteristics such as tilt and rotation while maintaining low cost~\cite{10.1145/3559400.3565595}. This technique has been widely applied in motion analysis for activities such as golf~\cite{king2008wireless, nam2013golf, fitzpatrick2010validation} and archery~\cite{phang2024archery, zhao2016archery}, where the tools are structurally simple but require precise motion tracking. Although some errors exist~\cite{fedorov2015using}, they can be effectively mitigated through algorithmic compensation~\cite{albaghdadi2019optimized, 10724704, 9289769, 10.1007/978-981-13-2553-3_36} enhancing the adaptability of the approach. Our findings demonstrate that the collected data sufficiently reveal trends in brush tilt and rotation and can be used to compare students’ and instructors’ writing. Thus, considering the balance of research needs, device availability, flexibility, and cost-effectiveness, we adopted this method.

In addition to motion capture technologies, accurately capturing the written characters themselves is equally important in calligraphy studies. To address this, we adopted top-down cameras to record the writing process, which aligns with the intuitive way humans observe calligraphy results. Furthermore, for detecting the lifting motion of the brush, we chose not to rely on the previously adopted inertial sensors. This is because variations in writing pressure during strokes make it challenging to define a clear boundary between writing and non-writing heights. Instead, we used a side-view camera (a smartphone) to directly detect contact between the brush tip and the paper. This combination of top-down and side-view visual capture ensures comprehensive data collection while preserving the natural flow of the writing process.

\subsection{Multilevel Support for Calligraphy Learning}

CalliSense offers a layered scaffolding approach tailored to the needs of learners at different stages, grounded in the imitation-centric nature of calligraphy learning. From beginner to advanced levels, learners progress by first deeply understanding the detailed brushwork of exemplary models and gradually integrating personal styles to achieve creative expression.

For beginners, the system clearly visualizes the intricate details that are often challenging to grasp, helping them identify key elements of calligraphy from the start. By breaking down techniques into manageable steps, learners can build a solid foundation as they gradually comprehend the complexity of brushstrokes. For advanced learners, CalliSense provides more sophisticated support. As they develop a foundational understanding of brushwork and rhythm, the system delivers detailed process data—such as brush rotation and stroke angles—that enables deeper analysis of these intricate movements, refining their technique and comprehension. Even for those who have mastered all brushwork techniques, exploring and appreciating the writing process behind others’ creations can provide valuable inspiration.

Moreover, self-reflection on brushwork details is an indispensable part of practice. Writing involves multiple focal points, making it difficult for learners to thoroughly review every action without assistance. CalliSense ensures traceability of the writing process, allowing learners to pause, replay key movements, and save practice records for future analysis and improvement. By dynamically deconstructing each stroke, the system helps learners clearly understand how multiple actions combine to form complete brushstrokes, significantly enhancing the effectiveness of kinesthetic learning.

\subsection{Integration of Independent Learning and Teacher-Guided Scenarios}
\textbf{Guided and self-directed learning.} In terms of application scenarios, the primary design goal of CalliSense is to support teacher-guided learning while holding potential for independent use as a self-learning tool. On one hand, the visualization of calligraphy processes aids teachers in explaining techniques. On the other hand, students can engage in autonomous practice by comparing their writing data with pre-saved examples from their instructors. However, despite efforts to ensure the system's hardware and software adaptability to both scenarios, CalliSense's current capabilities are not yet sufficient to fully support independent learning, particularly for beginners.
User interviews revealed challenges in standardizing calligraphy evaluation. Students often struggle to discern which differences in their writing require correction and which are acceptable when comparing their data with that of their teacher. Additionally, the complexity of calligraphy skills makes it difficult for learners to identify their current focus without explicit guidance. Supporting self-learning through CalliSense could involve designing features such as guided learning sequences~\cite{brydges2010comparing, duschl2011learning} or gamification to enhance engagement~\cite{suh2018enhancing, mohamad2018gamification}.

\textbf{Post-class learning tool.} Although CalliSense is currently centered on classroom support, its role in post-class review warrants further exploration. In this scenario, students—having identified key areas to improve—can independently experiment. Beyond mimicking the teacher’s techniques, they may explore variations in brush techniques and observe their impact on line quality. Repeated practice can lead to unintentional moments of success; for example, a student may accidentally replicate a line with texture strikingly similar to that of their teacher. These serendipitous moments, often fleeting, can be captured and analyzed using CalliSense’s ability to record detailed brushstroke parameters.

\textbf{ Peer comparison.} The system’s comparison functionality could also extend to peer interactions, where ``incorrect demonstrations'' play a vital role in learning~\cite{BOOTH201324}. By comparing brushstroke samples among classmates, students could deepen their understanding of calligraphy techniques.

\subsection{Diverse Applications Beyond the Classroom}
While initially designed for calligraphy teaching, the system's applications extend far beyond the classroom. By capturing fine dynamic details and visualizing data, it unveils hidden techniques in traditional calligraphy, surpassing the limitations of static demonstrations and addressing a long-standing challenge in calligraphy research: quantifying stroke effects and formation mechanisms~\cite{shi2023aesthetics}. We focused on teaching because of the high information transmission requirements, making it ideal for simplifying complex techniques. With parameters covering key variables like brush position, rotation, force, and speed, the system is well-suited for creation analysis, calligraphy exchanges, and academic discussions, showcasing its adaptability across a range of scenarios.

\subsection{Process Data Collection and Cross-Disciplinary Applications}
This study introduces a new method for collecting data on the calligraphy creation process, \REVISE{which is closely related to the concept of externalizing tacit knowledge\cite{ahmad2011influence, Virtanen_2011}}. While Chinese calligraphy emphasizes the final result, learners must understand how to achieve optimal stroke quality through brush posture, grip strength, and writing speed. \REVISE{These aspects of knowledge are often difficult to articulate or teach, as they are deeply embedded in the calligrapher’s muscle memory and sensory experience. Our system visualizes and analyzes these aspects of the creation process, helping learners connect hand control with stroke expression and providing precise feedback to facilitate skill mastery. By making these implicit elements accessible and comprehensible, our approach not only bridges the gap between expert practice and novice learning but also preserves and transmits the intricate art of calligraphy in a more structured and teachable form.}

This method also applies to other fields requiring fine motor control. A similar application is painting, which, like calligraphy, relies on brush control to affect stroke quality. Other areas include sports like golf, where posture recognition is crucial, and musical instrument performance, such as playing the violin. Additionally, by capturing brushstroke and hand force data, virtual pens in virtual reality (VR) and augmented reality (AR) can enhance learning experiences and support remote teaching.


\subsection{Feasibility of Crowdsourced Data Collection}
The low cost of the system's equipment makes it accessible for widespread use, enabling large-scale data collection through crowdsourcing. Calligraphy enthusiasts, researchers, and students can contribute to data collection using simple devices, helping build a diverse calligraphy dataset. This approach can gather samples from various countries and regions, capturing writing habits across different skill levels and forming a representative dataset. By showcasing the dynamic parameters of brush movements throughout the writing process, this dataset addresses a gap in previous ICH research, particularly the often-overlooked factor of hand force. This process-based data collection not only offers new tools for calligraphy preservation but also provides significant technical methods for broader ICH research.

\subsection{Limitations and Future Work}
This study has limitations in capturing fine brush details in small characters (such as Xiaokai), as the strokes are too small to accurately record subtle variations. However, the brush techniques for small characters can be transferred from large character practice, making the system still valuable in large character training. Currently, the system relies on teacher explanations to interpret complex brush parameters. In the future, large language models (LLMs) could be introduced to automatically recognize and interpret these parameters, reducing dependence on teachers and enhancing system autonomy.

In terms of image recognition, overlapping strokes can sometimes cause errors in complex lighting or reflective conditions. Future research should focus on optimizing skeletonization and time alignment algorithms to improve system adaptability in these challenging environments. Future studies could also explore ink dynamics\cite{Matsumaru_2017jaciii}, investigating the relationship between ink density and brushstrokes, and generating dynamic textures based on parameter mapping. Additionally, developing real-time feedback that combines touch, sound, and visual cues~\cite{10.1145/3281505.3281604, 10.1145/3305367.3327993} could help learners instantly correct mistakes during practice. The system could further support long-term progress analysis, tracking and visualizing learners' brush movements over time to assist teachers and students in evaluating progress. In remote teaching, comparing student data with standard datasets would enable personalized feedback, improving the accuracy of guidance.

Mastering complex brush techniques often requires long-term practice. While this study focuses on the short-term effectiveness of CalliSense in enhancing brushwork learning, mastering short-term skills may provide a foundation for long-term skill consolidation and transfer~\cite{billing2007teaching}. Additionally, during the user research phase, we observed that clarifying key calligraphy brushwork concepts within the CalliSense curriculum significantly facilitates subsequent repetitive practice. This approach aligns closely with the principles of Scaffolding Theory~\cite{van2002scaffolding}. Future research could design longitudinal follow-ups to evaluate students' skill retention and application over time. For instance, tracking students' handwriting performance one month or longer after using the system, as well as assessing whether they can flexibly apply learned techniques across different calligraphy styles. Additionally, building on the positive findings from current student feedback, future studies could explore how the system might support self-directed long-term learning. By observing learning outcomes over an extended period, further evidence could be gathered on the role of clear visual feedback in fostering sustained progress in calligraphy learning.

\section{Conclusion}
This paper introduces CalliSense, an interactive educational tool designed to support Chinese calligraphy brushstroke techniques through process-based learning. Semi-structured studies revealed that brushstroke details are often difficult to detect, especially for beginners who tend to overlook these techniques. To address this, we developed a comprehensive solution that utilizes low-cost devices, such as smartphones, pressure sensors, and inertial sensors, to capture key parameters during the writing process. These parameters are aligned with the written characters and visualized through an intuitive interface. 

User studies demonstrate that CalliSense significantly enhances students' understanding of critical brushstroke techniques compared to traditional teaching methods, reducing their neglect of important details and strengthening their brushstroke awareness. Additionally, the system provides teachers with an effective tool for communicating complex writing actions, improving instructional efficiency. Overall, our work presents an innovative approach to calligraphy education, not only enhancing learners' awareness and execution of brushstroke techniques but also offering technological support for the preservation of intangible cultural heritage. Future work will focus on further optimizing the system and exploring its applications in broader cultural and educational contexts.

\begin{acks}
This research was partially supported by the National Natural Science Foundation of China (No. 62202217), Guangdong Basic and Applied Basic Research Foundation (No. 2023A1515012889), and Guangdong Key Program (No. 2021QN02X794). We thank all of our study participants for their insightful discussions and feedback. We acknowledge the partial use of a large language model (LLM), specifically ChatGPT, to assist in the writing process. The LLM was employed as a tool for polishing the manuscript to enhance the clarity and quality of the text.
\end{acks}

\bibliographystyle{ACM-Reference-Format}
\bibliography{sample}





\end{document}